\documentstyle[12pt,aaspp4]{article}
\tightenlines

\font\boldsym=cmmib10

\def	\amu	{\,{\rm amu}}
\def	\Angstrom	{\,{\rm \AA}}		% Angstrom

\def	\bepsilon {{\hbox{\boldsym\char'017}}}	%bold \epsilon

\def 	\bE	{{\bf E}}
\def	\beq	{\begin{equation}}

\def    \bmu    {{\hbox{\boldsym\char'026}}}	%bold \mu

\def    \bomega {{\hbox{\boldsym\char'041}}}	%bold \omega

\def	\bv	{{\bf v}}

\def	\C	{{\rm C}}
\def	\calF	{{\cal F}}
\def	\cm	{\,{\rm cm}}

\def	\D	{{\rm D}}
\def	\debye	{\,{\rm debye}}

\def	\eeq	{\end{equation}}

\def	\erg	{\,{\rm ergs}}
\def	\eV	{\,{\rm eV}\,}
\def	\g	{\,{\rm g}}

\def	\GHz	{\,{\rm GHz}}
\def	\gtsim	{\simgt}
\def	\H	{{\rm H}}
\def	\He	{{\rm He}}
\def	\HH	{{\rm H}_2}
\def	\Hz	{\,{\rm Hz}}
\def	\IR	{{\rm IR}}

\def	\ISRF	{{\rm ISRF}}
\def	\Jy	{\,{\rm Jy}}
\def	\K	{{\rm K}}

\def	\ltsim	{\simlt}
\def	\micron	{\mu{\rm m}}
\def	\MRN	{{\rm MRN}}

\def	\nH	{n_{\rm H}}

\def	\s	{\,{\rm s}}
\def	\sr	{\,{\rm sr}}

\def	\ted	{{\tau_{ed}}}
\def	\tH	{{\tau_\H}}

\def    \simlt  {\lower.5ex\hbox{$\; \buildrel < \over \sim \;$}}
\def    \simgt  {\lower.5ex\hbox{$\; \buildrel > \over \sim \;$}}

\lefthead{Draine \& Lazarian}
\righthead{Grain Radiation}

\begin{document}

\begin{center}
%{\bf *** DRAFT \today\  *** Please do not circulate.}
{Submitted to {\it The Astrophysical Journal}}
\end{center}

\title{Electric Dipole Radiation from Spinning Dust Grains}

\author{B.T. Draine \& A. Lazarian}
\affil{Princeton University Observatory, Peyton Hall, Princeton,
NJ 08544}

% enable following line to print table of contents
%\tableofcontents

\begin{abstract}

We discuss the 
rotational excitation of small interstellar grains and the resulting 
electric dipole radiation from spinning dust.
Attention is given to excitation and damping
of grain rotation by: 
collisions with neutrals; 
collisions with ions;
``plasma drag'';
emission of infrared radiation; 
emission of electric dipole radiation;
photoelectric emission;
and formation of $\HH$ on the grain surface.
Electrostatic ``focussing'' can substantially enhance the rate of
rotational excitation of grains colliding with ions.
Under some conditions, ``plasma drag'' 
-- due to interaction of the electric dipole moment of
the grain with the electric field produced by passing ions --
dominates both rotational
damping and rotational excitation.
We introduce dimensionless functions $F$ and $G$
which allow direct comparison of the contributions of different
mechanisms to rotational drag and excitation.

Emissivities are estimated for dust in 
different phases of the interstellar medium, including
diffuse HI clouds, warm HI, low-density photoionized gas, and
cold molecular gas.
Spinning dust 
grains could explain much, and perhaps all, of 
the 14 - 50 GHz background component 
recently observed by 
Kogut et al. (1996), 
de Oliveira-Costa et al. (1997)
and Leitch et al. (1997).
Future sensitive measurements of angular structure in the microwave
sky brightness from the ground and from space should detect this emission
from high-latitude HI clouds.
It should be possible to detect rotational emission from small
grains by ground-based pointed observations
of molecular clouds.

\end{abstract}


\section{Introduction}

Experiments to map the cosmic 
microwave background radiation have stimulated renewed
interest in diffuse Galactic emission. Most
recently, 
Kogut et al (1996), 
de Oliveira-Costa et al. (1997) and Leitch et al. (1997) 
have reported a new component of galactic microwave
emission which is correlated with $100\micron$ thermal
emission from interstellar dust.

Kogut et al. found the emission excess to have 
$I_\nu(31.5{\rm GHz})\approx I_\nu(53{\rm GHz})$, and
Leitch et al. found 
$I_\nu(14.5{\rm GHz})\approx I_\nu(32{\rm GHz})$,
consistent with the spectrum of free-free emission, but nondetection of
H$\alpha$ emission in these directions 
is inconsistent with free-free emission accounting for
the microwave excess unless the plasma temperature $T\gtsim 10^6\K$.
Leitch et al. therefore 
proposed that the observed emission was free-free emission from shock-heated
gas in a supernova remnant.

Draine \& Lazarian (1998, hereafter DL98) 
showed, however, that the observed microwave
excess could not be due to free-free emission from hot gas, as this
would require an energy injection 
rate at least 2 orders of magnitude greater than
that the energy input due to supernovae.
DL98 showed that the microwave excess could in fact be electric
dipole emission from rapidly-rotating dust grains.

To predict the intensity of dipole emission one needs 
(1) the numbers of small grains, 
(2) their dipole moments, and 
(3) their rotational velocities.
The observed intensity of 12 and $25\micron$ emission from interstellar
clouds allows us to estimate the numbers of very small grains
(Leger \& Puget 1984; Draine \& Anderson 1985; Desert, Boulanger,
\& Puget 1990).
To estimate the dipole moments, we consider the likely displacements
between the
charge and mass centroids for grains, plus the intrinsic dipole
moment arising from polarized chemical bonds in the grain material.
However, the main thrust of the present paper is a comprehensive study
of the rotational dynamics of small grains, in order to estimate
their rotation rates.

Ferrara \& Dettmar (1994) showed that small grains rotating with
thermal rotation rates could produce detectable radio emission, but
they did not address the details of rotational excitation and
damping.
Rotational excitation of small grains (= large molecules)
has been discussed previously by
Rouan et al (1992), who described the effects of collisions
with gas atoms and absorption and emission of radiation.
In the present work we reexamine this problem, and include the important
effects of collisions with ions, which were neglected in the study by
Rouan et al.
We also include the effects of ``plasma drag'' due to interaction 
of the grain with passing
ions, first considered by Anderson \& Watson (1993).
We derive rates for rotational damping and excitation due to plasma drag
which are somewhat larger than estimated by
Anderson \& Watson (1993).

We discuss rotational excitation as a 
function of both grain size and environmental conditions.
Electrostatic focussing of ions makes them very effective at
delivering angular momentum to the grains; ion collisions can
dominate the rotational excitation of very small grains even in
predominantly neutral regions.

For our adopted population of very small grains,
we predict the microwave
emissivity of various phases of the interstellar medium, ranging from
diffuse gas to molecular clouds.
We expect detectable levels of emission from spinning dust grains in 
all of these phases.

The paper is organized as follows:
After describing the environments which we consider (\S\ref{sec:media}),
we discuss the electric dipole moments expected for
small dust grains (\S\ref{sec:grain_properties}).
The various rotational damping processes are
reviewed in \S\ref{sec:rotational_damping},
and the 
rotational excitation mechanisms in \S\ref{sec:rotational_excitation}.
Using these rates for rotational excitation and damping, we calculate
the resulting rate of grain rotation in \S\ref{sec:rot_rate}.
The importance of the impulsive nature of the rotational excitation
is discussed in 
\S\ref{sec:impulsive_J}, and the effects of centrifugal
stresses in \S\ref{sec:centrifugal}.

The reader interested primarily in the predicted emission may wish to skip
\S\S\ref{sec:media}--\ref{sec:centrifugal} and proceed directly to
\S\ref{sec:sized}, where we describe our assumptions concerning the
grain size distribution in various environments.
We present the microwave emission spectra expected for grains in both
diffuse regions and molecular clouds in \S\ref{sec:emissivity}.
The detection of this microwave emission from pointed observations
of dense clouds is discussed in \S\ref{sec:darkcloud}.
The principal uncertainties in our
estimates are discussed in \S\ref{sec:discussion}, and
our results are summarized in \S\ref{sec:summary}.

\section{Environments \label{sec:media}}

We will consider grains in five different idealized ``phases'' of the
interstellar medium: 
``Cold Neutral Medium'' (CNM),
``Warm Neutral Medium'' (WNM),
``Warm Ionized Medium'' (WIM),
``Molecular Cloud'' (MC),
and
``Dark Cloud'' (DC).
In Table \ref{tab:phases} we give the 
adopted values of $n_\H$ (H nucleon density),
$T$ (gas kinetic temperature),
$\chi$ (starlight intensity relative to the average starlight background),
and other properties of these phases.

\section{Grain Properties\label{sec:grain_properties}}
\subsection{Grain Sizes and Shapes}

We will characterize the grain size by the ``volume equivalent'' radius 
$a\equiv 10^{-7}a_{-7}\cm$, the radius
of a sphere of equal volume.
We will assume
a density
$\rho\approx2.\g\cm^{-3}$,
having in mind a carbonaceous material with C:H$\approx$3:1;
for C:H::3:1 the mass per atom is $m\approx9.25\amu$, and
the number of atoms in the grain is then
\beq
N = 545 a_{-7}^3 ~~~.
\eeq
We will consider grains down to a minimum size $N\approx25$, or
$a_{-7}\approx0.36$.

We require the moment of inertia as a
function of grain size, so assumptions must be made concerning the
grain shape.
The smallest grains may be chainlike, as favored by Thaddeus (1995),
or sheetlike, as expected for polycyclic aromatic hydrocarbons.
For generality, we will suppose that grains with
$a < a_1$ are cylindrical with diameter $d$,
grains with $a_1 < a < a_2$ are disklike with thickness $d$,
and grains with $a > a_2$ are spherical;
we take
$d=3.35\times10^{-8}\cm$, the interlayer separation in graphite.
The actual geometry of small grains is uncertain.
Here we usually take
$a_1=0$ and 
$a_2=6\times10^{-8}\cm$ (grains with $N<120$ atoms are disklike),
but we will also consider the possibility that 
$a_1=4\times10^{-8}\cm$ (grains with $N<50$ atoms are linear).

The physics of rotational excitation and damping will be explicitly
derived for spherical grains; when nonspherical grains are considered,
we will attempt to use results for spherical grains but with the radius
$a$ for the sphere replaced by an appropriate length scale for the
nonspherical shape.
For convex shapes, the rates for collisions of neutral grains with atoms 
depends on the surface area $S$, and we therefore define the 
``surface-equivalent'' radius $a_s$,
the radius
of a sphere with equal surface area:
\beq
4\pi a_s^2\equiv S ~~~.
\label{eq:asdef}
\eeq
The excitation of rotational kinetic energy by collisions 
is proportional to
$\int r^2 dS$, where $r$ is the 
distance from the surface point to the center-of-mass.
We therefore
define the ``excitation equivalent'' radius
$a_x$, the radius of a sphere
with equal $\int r^2 dS$:
\beq
4\pi a_x^4 \equiv \int r^2 dS ~~~.
\label{eq:axdef}
\eeq
Another important quantity is the ratio of the moment of inertia to the
value for a sphere of radius $a$:
$\xi \equiv I / (0.4 M a^2)$, where $M$ is the grain mass and $I$ is the
largest eigenvalue of the moment of inertia tensor.
\subsection{Grain Charge}

Grains acquire charge through photoelectric emission and collisions
with electrons and ions.
Bakes \& Tielens (1994) have discussed 
the rate $\dot{N}_{pe}(Z)$ of photoelectron emission
from a carbonaceous particle with charge $Ze$;
we adopt their rates with minor modifications (Weingartner \& Draine 1998).
Collisional charging processes have been discussed by 
Draine \& Sutin (1987),
whose rates we employ here.
When considering nonspherical grains, we 
approximate the potential as $U\approx Ze/a_s$.
In computing the electron and ion capture rates, we assume a spherical
geometry with a radius $a_s$.

\begin{figure}
\epsscale{1.00}
\plotone{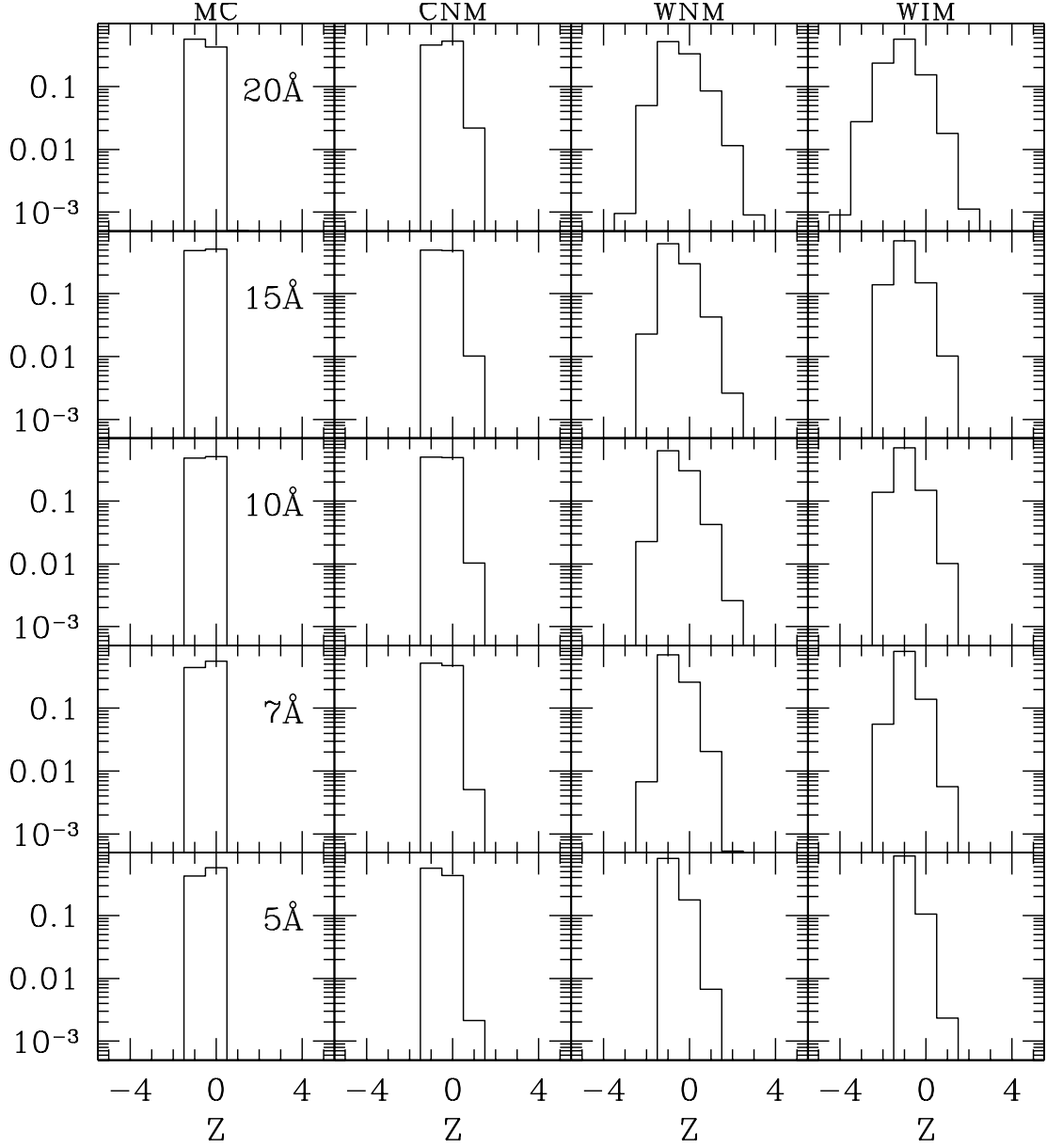}
\caption{
	\label{fig:zdists}
	Charge distribution functions for grains of radii
	$a=5,7,10,15,20\Angstrom$ for ``Molecular Cloud'' (MC),
	``Cold Neutral Medium'' (CNM),
	``Warm Neutral Medium'' (WNM), and
	``Warm Ionized Medium'' (WIM) conditions.
	}
\end{figure}

Let $\dot{N}_e(Z)$ and $\dot{N}_i(Z)$ be the
rate of electron capture and ion impact for
a grain of radius $a_s$ and charge $Ze$.
With these rates, we solve the usual equations to obtain the steady-state
charge distribution function $f(Z)$ for grains of radius $a$:
\beq
\left[\dot{N}_i(Z)+\dot{N}_{pe}(Z)\right]f(Z)
=\dot{N}_e(Z+1)f(Z+1) ~~~.
\eeq
Selected charge distribution functions $f(Z)$ are shown in 
Figure \ref{fig:zdists} for CNM and WNM conditions.

\begin{figure}
\epsscale{1.00}
\plotone{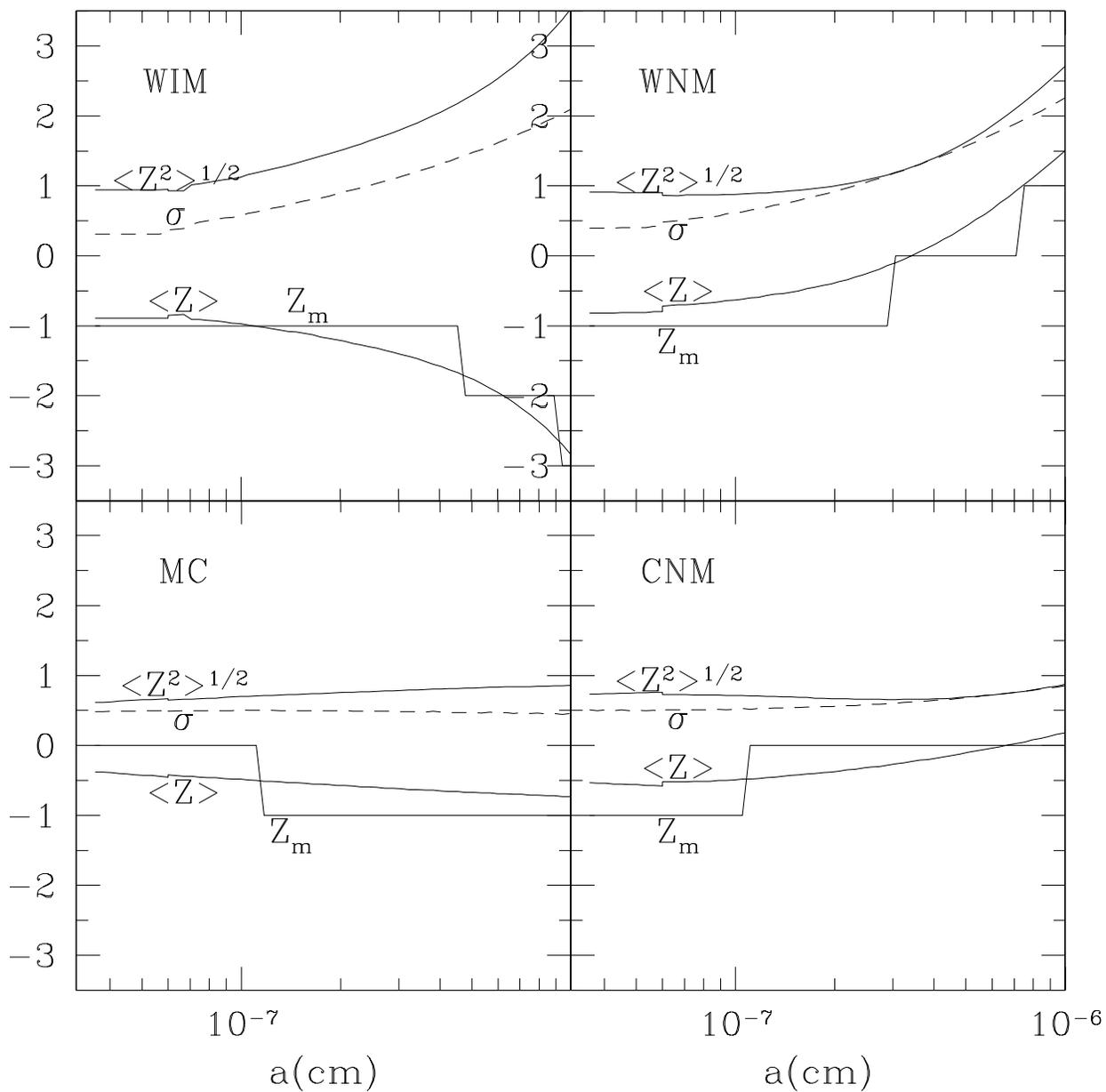}
\caption{
	\label{fig:zbar_etc}
	Mode $Z_m$, centroid $\langle Z\rangle$, 
	rms charge $\langle Z^2\rangle^{1/2}$, and standard deviation
	$\sigma_Z$ for grains of radius $a$ for CNM and WNM conditions.
	}
\end{figure}

The charge distribution for grains of radius $a$
has mode $Z_m$,
centroid $\langle Z\rangle$, and variance 
$\sigma_Z^2=\langle Z^2\rangle-\langle Z\rangle^2$; these are shown in
Figure \ref{fig:zbar_etc}.
The time scale for changes in the grain charge is of interest.
We define a characteristic charging time by
\beq
\tau_Z = 
{1+\sigma_Z^2\over \dot{N}_e(Z_m)+\dot{N}_i(Z_m)+\dot{N}_{pe}(Z_m)}
\label{eq:tauz}
\eeq
When $\sigma_Z^2\ll1$, $\tau_Z$ is simply the lifetime of charge state
$Z_m$; when $\sigma_Z^2\gg1$, $\tau_Z$ is an estimate for the time
required for the grain charge to change by of order the
standard deviation $\sigma_Z$ of the charge distribution.

\begin{figure}
\epsscale{1.00}
\plotone{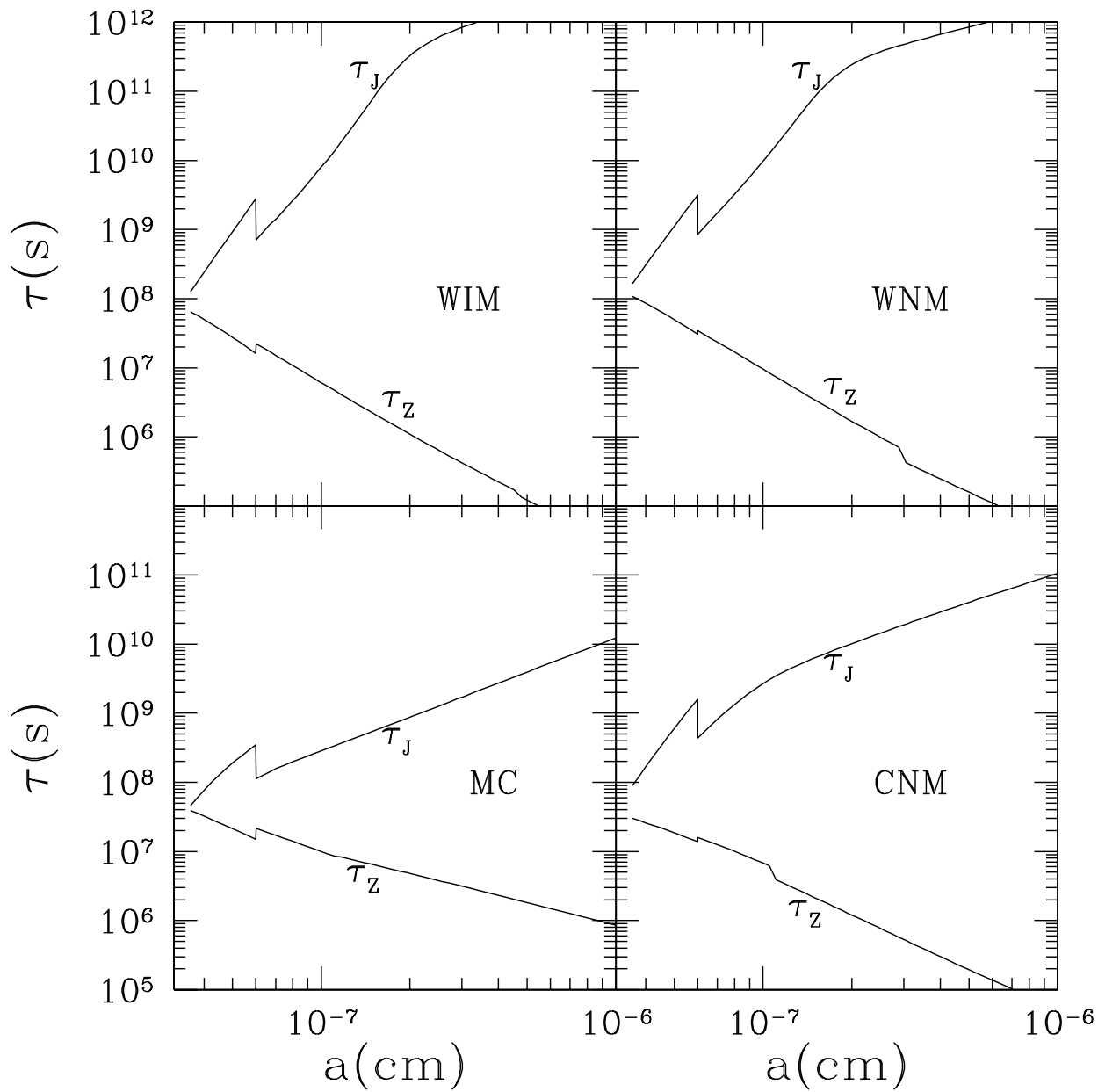}
\caption{
	\label{fig:tauZ_and_tauJ}
	Characteristic time scale $\tau_Z$
	[see eq. (\protect{\ref{eq:tauz}})] for changes in the grain charge $Ze$.
	Also shown is the characteristic rotational damping time $\tau_J$
	[see eq. (\protect{\ref{eq:tauJ}})] for a grain with charge $Z_me$.
	It is apparent that the approximation $\tau_Z \ll \tau_J$ is
	excellent for all except the smallest ($a < 4\Angstrom$) grains.
	}
\end{figure}

Figure \ref{fig:tauZ_and_tauJ} shows $\tau_Z$ as a function of
grain size $a$ for CNM and WNM conditions.

\subsection{Quantum vs. Classical Dynamics\label{sec:QvsC}}

The grain angular momentum $I\omega$ is quantized.
If the rotational kinetic energy is $(3/2)kT_{rot}$, then the
rotational quantum number
\beq
J = {I\omega\over \hbar} =
5.85 \xi^{1/2} N^{5/6}\left({T_{rot}\over 100\K}\right)^{1/2} ~~~.
\eeq
The smallest grain we consider will have $\xi^{3/5}N\gtsim 35$, so that
$J \gtsim 113(T_{rot}/100\K)^{1/2} \gg 1$.
We are therefore justified in using classical mechanics to discuss
the grain rotation.

\subsection{Grain Dipole Moments\label{sec:dip_moments}}

A spinning grain
radiates power
\beq
P=\frac{2}{3}\frac{\omega^4\mu^2\sin^2\theta}{c^3}~~~,
\eeq
where $\theta$ is the angle between the angular velocity $\bomega$
and electric dipole moment $\bmu$. We assume that the grain dipole
moment $\bmu$ can be written
\beq
{\bmu}={\bmu}_i + {\bepsilon}Zea_x~~~,
\eeq
where $\bmu_i$ is the intrinsic dipole moment of an uncharged grain,
$Ze$ is the grain charge, and
the vector ${\bepsilon}a_x$ is
the displacement between the grain center of mass and the charge centroid
[the length $a_x$ is defined by eq.(\ref{eq:axdef})].
As noted by Purcell (1976), an irregular
grain, even if perfectly conducting, will, in 
general, have its charge centroid displaced from the mass centroid. The 
magnitude of the displacement depends on the grain shape;
we will assume it to be $\sim 10\%$ of $a_x$,
or $|{\bepsilon}|\approx 0.1$.  Thus
\beq
\epsilon Zea_x = 4.8 \left({a_x\over10^{-7}\cm}\right)
Z \left({|\bepsilon|\over0.1}\right) \debye ~~~.
\eeq

Neutral molecules can of course have electric dipole moments.
Except for the case of highly symmetric molecules, these dipole moments
are expected to be appreciable.
If the grain material has short range order, but minimal long-range order,
then it consists of more-or-less randomly-arranged
chemical substructures, and
\beq
\mu_i \approx N^{1/2}\beta ~~~.
\label{eq:mu_i}
\eeq
To estimate the likely magnitude of $\beta$, we show in Table \ref{tab:bonds}
the dipole moments associated with particular bonds in hydrocarbon molecules.
Dipole moments of $\sim1\debye$ are evidently typical in hydrocarbon
molecules.
Of course, a highly symmetric molecule would
have a net dipole moment of zero, but we do not expect such perfect symmetry
to be the norm for large interstellar molecules.
The larger molecules should have minimal long-range order, having
grown by random arrival of gas atoms or by coagulation with other small
molecules.
It might be argued that 
very small molecules will be predominantly symmetric
because only the most robust molecules will survive in the interstellar 
ultraviolet radiation field.
For example, 
the highly symmetric polycyclic aromatic hydrocarbon $\C_{24}\H_{12}$ 
coronene (with $\mu_i=0$)
has been considered as a candidate for the very smallest
interstellar grains.
We note, however, that under interstellar conditions the carbon skeleton for
very small grains
is expected not to be fully hydrogenated (Omont 1986).
Such radicals are not easily studied in the laboratory,
and dipole moments are unavailable.
We can estimate, however, that substitution of OH for one of the H in
coronene would produce a molecule with a dipole moment of $\sim1.6\debye$,
or $\beta\approx0.3\debye$.
In Table \ref{tab:betas} we give electric dipole moments for 
selected molecules, as well as the corresponding values of 
$\beta=\mu_i/N^{0.5}$.
%Carbon chains, suggested by Thaddeus (1995) as the carriers of the diffuse
%bands, tend to have larger dipole moments for a given $N$ (see, e.g.
%cyanopolyyne $\H\C_{11}\N$ in Table \ref{tab:betas}).

Based on Table \ref{tab:betas}, and the expectation that radicals will 
typically have
larger dipole moments than complete molecules, we
will take $\beta_0=0.4\debye$ as representative.\footnote{
	Rouan et al. (1992) assumed $\mu=3\debye$
	for $N=90$, or $\beta=0.32\debye$.
	}

In an ensemble of grains of radius $a$, we may suppose the direction of
$\bepsilon$ 
and 
the direction of the grain's intrinsic dipole moment $\bmu_i$
to be uncorrelated.
Thus, with $\epsilon=0.1$ and $\beta=0.4\debye$ 
the mean square dipole moment will be
\beq
\mu^2 = 
[(4.8)^2\left({a_x\over a}\right)^2 \langle Z^2\rangle + 
(9.3)^2a_{-7}]a_{-7}^2 (\debye)^2 ~~~.
\label{eq:mu2}
\eeq
If the orientation of $\bmu$ is uncorrelated with the angular
velocity $\bomega$, then 
$\langle\sin^2\theta\rangle=2/3$, 
and the expected radiated power per grain is just
\beq
P = {4\over9} {\mu^2\omega^4\over c^3} ~~~.
\label{eq:Pspin}
\eeq

\section{Rotational Damping\label{sec:rotational_damping}}

Gas-grain interactions, plasma-grain interactions, infrared emission,
and radio emission damp the rotation of small grains.
In this section we present the rates for these processes.

We consider a spherical grain with temperature $T_d$ in gas of
temperature $T$, H nucleon density $n_\H=n(\H)+2n(\H_2)$, and with
$n(\He)=0.1n_\H$.

In thermal equilibrium at temperature $T$,
the grain would have an rms rotation frequency
\beq
{\omega_T\over 2\pi} = 
\langle\nu^2\rangle^{1/2}
\approx 5.60\times 10^9 a_{-7}^{-5/2}\xi^{-1/2}T_2^{1/2}~~~{\rm Hz}~,
\label{eq:nurms}
\eeq
where
$T_2=T/100\K$.
If the dust temperature $T_d\neq T$, or there are sources of rotational
excitation and
drag other than the gas (see below), then the grain rotation rate 
will differ from eq.(\ref{eq:nurms}).

Here we summarize the contributions of gas-grain interactions to
both rotational damping and excitation;
details may be found in Appendix \ref{app:rotexc}.
It is convenient to ``normalize'' the various drag processes to
the drag which would be produced by ``sticky'' collisions in a pure
H gas of density $\nH$: thus for drag process $j$ we define the dimensionless
quantity $F_j$ such that
the contribution of process $j$ to the drag torque is
\beq
\label{eq:fdef}
I\left({d\omega\over dt}\right)_j \equiv
- \left[ n_\H \left( {8 k T \over \pi m_\H}\right)^{1/2} 
{2\pi a_x^4 m_\H \over 3} \right] 
\omega F_j  ~~~.
\eeq
Neutral grains in a gas of pure H atoms would have $F=1$, with a
rotational damping time
\begin{eqnarray}
\tH &=&
{4\xi\rho a^5 \over 5 n_\H m_\H a_x^4}\left( {\pi m_\H \over 8kT} \right)^{1/2}
\\
&=&
3.30\times10^{10}\xi
\left({20 \cm^{-3} \over n_\H}\right) T_2^{-1/2}
a_{-7}
\left({a\over a_x}\right)^4 
\s ~~~,
\end{eqnarray}
where we have assumed $\rho=2\g\cm^{-3}$.
We will use $\tH$ as a fiducial time scale for the
different sources of rotational damping.
Thus the rotational damping time due to process $j$ is
\beq
\tau_j^{-1} = F_j ~ \tH^{-1} ~~~.
\eeq
The linear drag processes are additive:
\beq
F = F_{n} + F_{i} + F_p + F_\IR ~~~,
\label{eq:fsum}
\eeq
where $F_{n}$, $F_i$, $F_p$, and $F_\IR$ are the
contributions from neutral impacts,
ion impacts, 
plasma drag,
and
thermal emission of infrared photons.

\subsection{Collisional Drag}

Collisional damping is a well-studied
process (see Jones \& Spitzer 1967). 
We assume that when species (e.g., H, H$_2$, He)
arrive at the grain surface they temporarily
``stick'' and then are desorbed with a thermal velocity distribution
relative to the local (moving) grain surface
(the possible effects of H$_2$ formation are considered separately
in \S\ref{sec:systematic}).
The dimensionless factor $F_j$ for species $j$,
which would be unity for H atoms and
neutral grains, depends on the grain radius $a$, charge state
$Z_g$, and on the gas composition and temperature $T$.

\begin{figure}
\epsscale{1.00}
\plotone{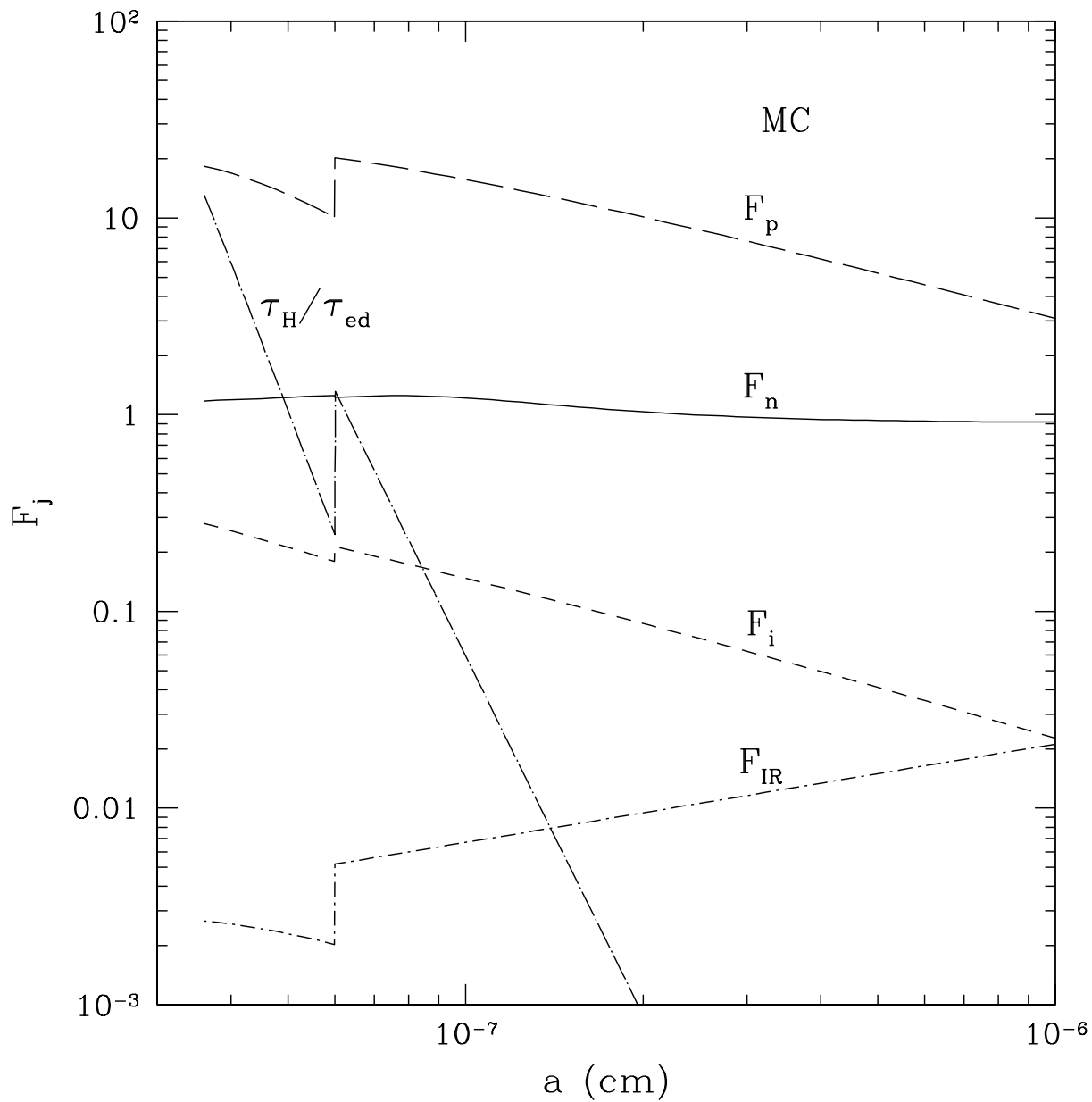}
\caption{
	\label{fig:F_MC}
	Dimensionless rotational drag functions for 
	neutral collisions [$F_n$, eq.(\protect{\ref{eq:F_n}})],
	ion collisions [$F_i$, eq.(\protect{\ref{eq:F_i}})], 
	plasma drag [$F_p$, eq.(\protect{\ref{eq:F_p}})],
	and infrared emission [$F_\IR$, 
	eq.(\protect{\ref{eq:Firc}}-\protect{\ref{eq:F_IR}})],
	for ``Molecular Cloud'' conditions.
	Plasma drag is more important than other 
	gas collisional processes.
	Also shown is the ratio $\tH/\ted$ showing
	the relative importance of electric dipole damping.
	For molecular cloud conditions, electric dipole damping
	is not important.
	}
\end{figure}

\begin{figure}
\epsscale{1.00}
\plotone{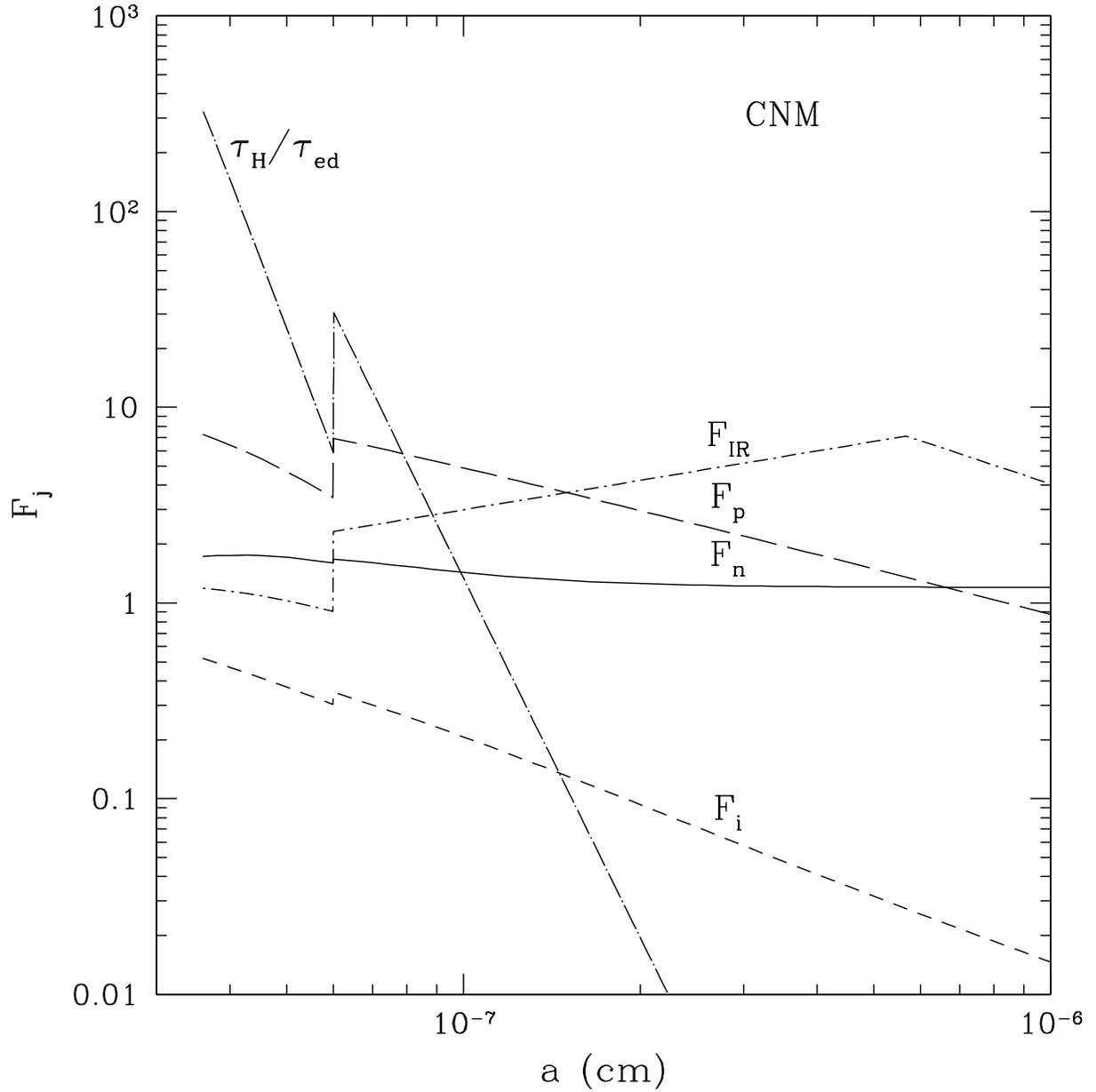}
\caption{
	\label{fig:F_CNM}
	Same as Fig. \protect{\ref{fig:F_MC}} but for ``Cold Neutral
	Medium'' conditions.
	Electric dipole damping dominates for $a\ltsim 4\times10^{-8}\cm$.
	Plasma drag $F_p$ is the dominant drag process for
	$4\times10^{-8}\ltsim a \ltsim 1.5\times10^{-7}\cm$.
	Infrared emission damping is dominant for $a\gtsim 1.5\times10^{-8}\cm$.
	}
\end{figure}

\begin{figure}
\epsscale{1.00}
\plotone{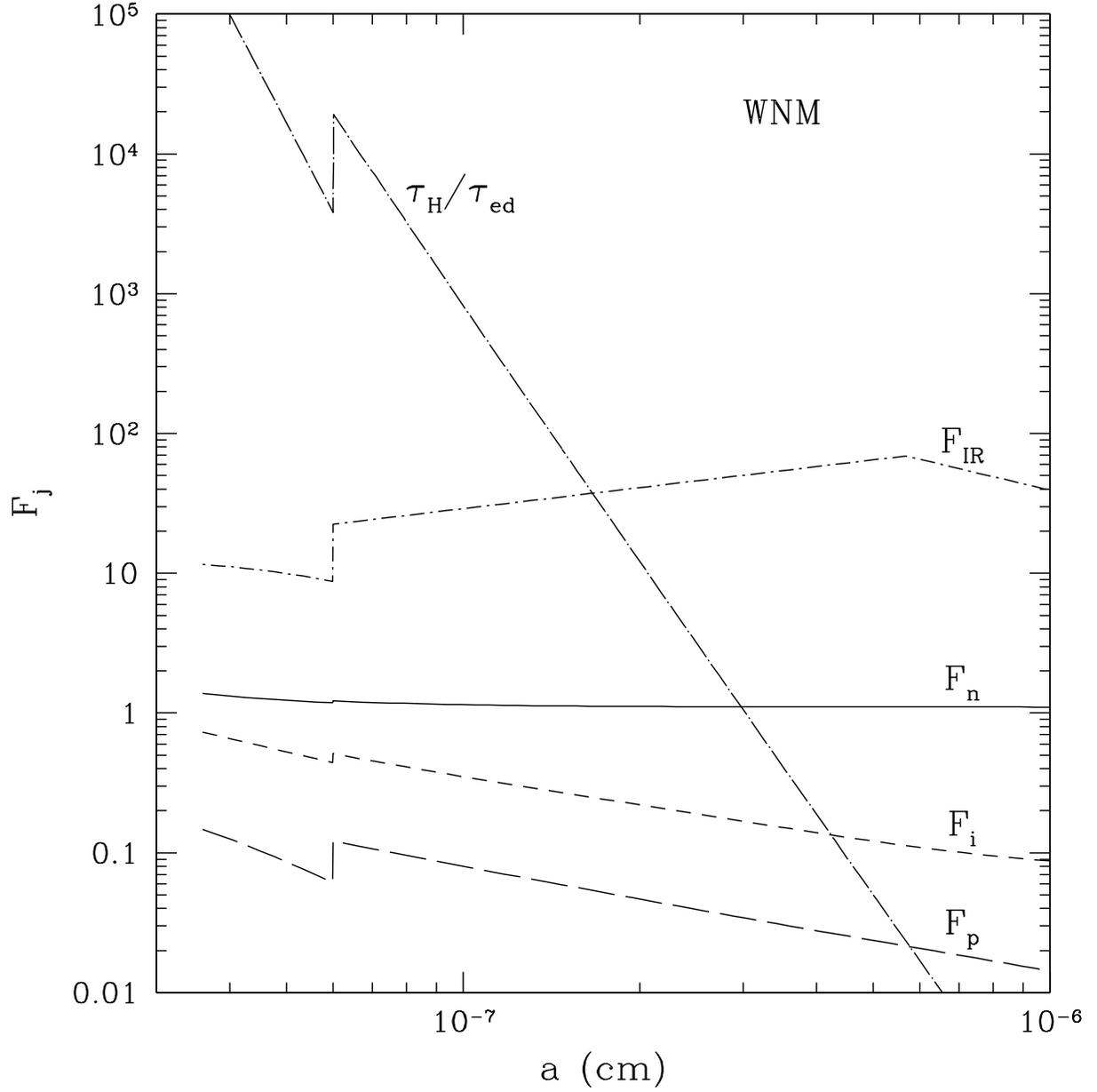}
\caption{
	\label{fig:F_WNM}
	Same as Fig. \protect{\ref{fig:F_MC}} but for ``Warm Neutral
	Medium'' conditions.
	Electric dipole damping dominates for $a\ltsim1.5\times10^{-8}\cm$,
	and IR damping dominates for larger grains.
	}
\end{figure}

\begin{figure}
\epsscale{1.00}
\plotone{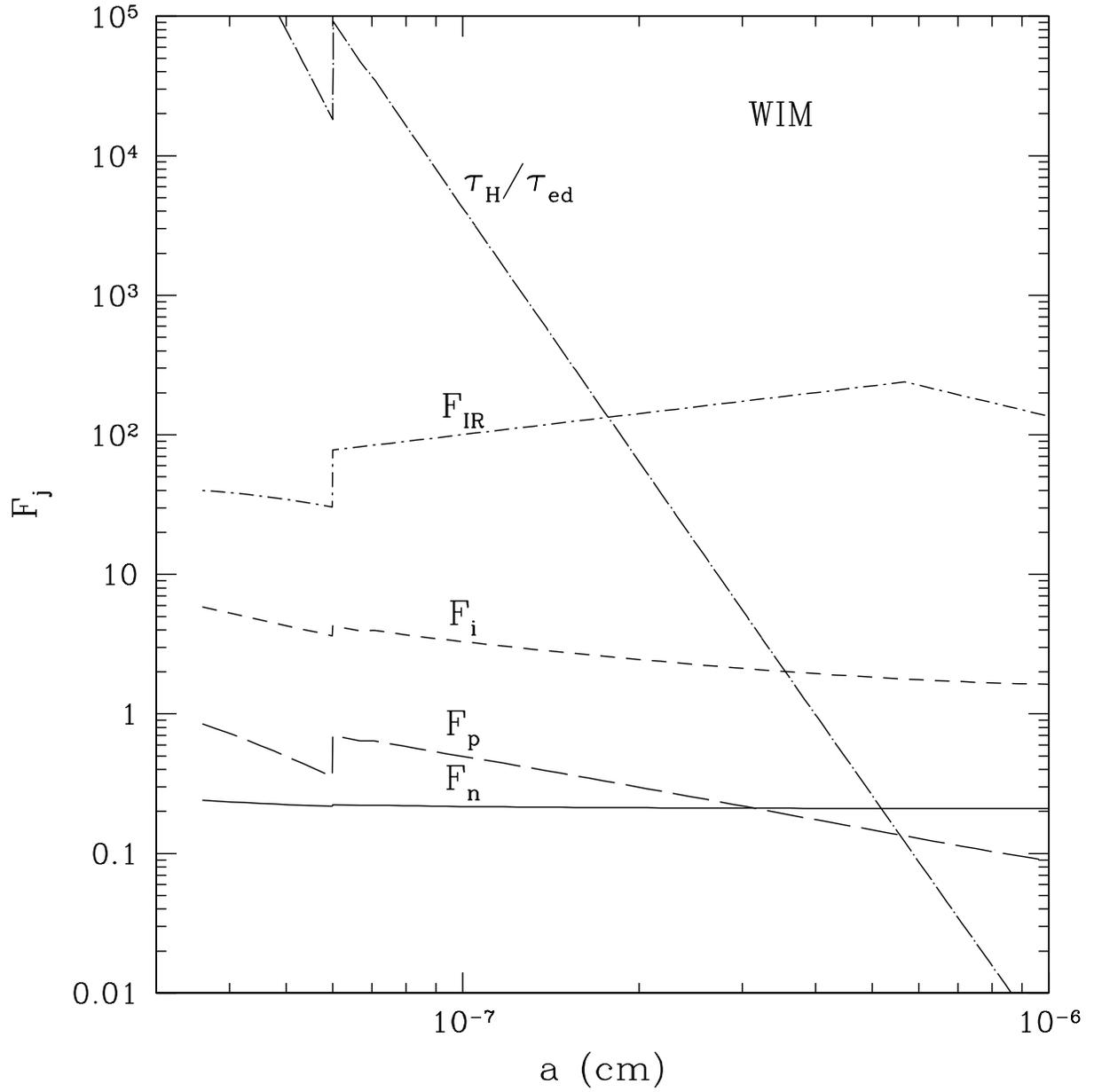}
\caption{
	\label{fig:F_WIM}
	Same as Fig. \protect{\ref{fig:F_MC}} but for ``Warm Ionized
	Medium''
	conditions.
	}
\end{figure}

Let $f(Z_g)$ be the probability that the grain will have a charge
$Z_ge$.
Because the charging timescale $\tau_Z$ is short compared to the timescale
$\tau_J$ over which the grain rotation rate changes appreciably, we can
average over the grain charge distribution $f(Z_g)$.
Then
(cf. Appendix \ref{app:F_n} and \ref{app:F_i})\footnote{
	For nonspherical grains we use the length scale
	$a_s$ to estimate the potential $\sim Ze/a_s$ and surface
	electric field $\sim Ze/a_s^2$.
	}
\begin{eqnarray}
F_{n}
&=&
\sum_{Z_g} f(Z_g) \sum_n {n_n \over n_\H}
\left({m_n \over m_\H}\right)^{1/2}
\left[ \exp(-Z_g^2\epsilon_n^2) + 
|Z_g|\epsilon_n \pi^{1/2} {\rm erf}(|Z_g|\epsilon_n)\right]
\label{eq:F_n} ~~~,\\
F_i&=&
\sum_{Z_g\neq0} f(Z_g) \sum_i 
{n_i \over n_\H}\left({m_i \over m_\H}\right)^{1/2} 
g_1\left( {Z_g Z_i e^2 \over a_s kT} \right)\nonumber\\
&&+ ~ f(Z_g=0)\sum_i {n_i \over n_H}\left({m_i\over m_\H}\right)^{1/2}
\left(1 + {\pi^{1/2}\over 2}Z_i\phi\right)
\label{eq:F_i}~~~,\\
\end{eqnarray}
where
\beq
g_1(x) = \left\{
\begin{array}{ll}
	1-x	& \mbox{if $x < 0$} \\
	e^{-x} & \mbox{if $x > 0$}
\end{array}
	\right. ~~~,
\eeq
\beq
\label{eq:epsilondef}
\epsilon_n^2 \equiv {e^2 \alpha_n \over 2 a_s^4 kT} ~~~,
\eeq
\beq
\label{eq:phidef}
\phi = \left({2 e^2\over a_s kT}\right)^{1/2} ~~~,
\eeq
$n_n$, $m_n$ and  $\alpha_n$ are the number density, mass, and 
polarizability of neutral species $n$.
Equation (\ref{eq:F_n}) requires summation over neutral species $n$,
and
eq. (\ref{eq:F_i}) over ion species $i$.
$F_n$ and $F_i$ are shown in Figs. \ref{fig:F_MC} -- \ref{fig:F_WIM}
for MC, CNM, WNM, and WIM conditions.

\subsection{Plasma Drag}

If the grain has an electric dipole moment $\mu$, then the interaction of
this electric dipole moment with passing ions couples the grain rotation
to the plasma (Anderson \& Watson 1993).
The contribution to the damping is derived in Appendix 
\ref{app:F_p}\footnote{
	In eq.(\ref{eq:F_p}) the factor $a_x^{-4}$ enters because of the
	normalization (\ref{eq:fdef}).
	}
\beq
F_{p} \approx 
{2e^2 \mu^2 \over 3a_x^4 (kT)^2}
\sum_i {n_i Z_i^2 \over n_\H}\left({m_i \over m_\H}\right)^{1/2} 
\left\{
\ln(b_\omega/a)+\cos^2\Psi\ln\left[\min(b_q,\lambda_{\rm D}/b_\omega)\right]
\right\}~~~,
\label{eq:F_p}
\eeq
where 
\beq
\lambda_{\rm D} \equiv \left( {kT \over 4\pi n_e e^2}\right)^{1/2}
\eeq
is the Debye shielding length,
\beq
b_\omega = (I/m)^{1/2} ~~~,
\eeq
\beq
b_q = I (2kT/m_i)^{1/2}/\hbar
\eeq
and $\Psi$ is the angle between $\bmu$ and $\bomega$.
Our result (\ref{eq:F_p}) differs from that of Anderson \& Watson (1993),
who neglect impact parameters with $b>b_\omega$.
In the present calculations we set $\cos^2\Psi=1/3$, appropriate if
$\bomega$ is randomly oriented relative to $\bmu$.\footnote{
	We note, however, that disklike grains would tend to have
	$\cos^2\Psi\approx1$, since internal relaxation will cause the
	principal axis of largest moment of inertia to align with the
	angular momentum, whereas $\bmu$ will tend to be perpendicular
	to the principal axis of largest moment of inertia.
	}
$F_p$ is shown in Figs. \ref{fig:F_MC} -- \ref{fig:F_WIM}.

\subsection{Infrared Emission}

A more subtle process is damping due to infrared emission
(Purcell \& Spitzer 1971; Martin 1972).
As shown in Appendix \ref{app:IRdamping},
emission of a thermally-excited 
photon of energy $h\nu$ from a spherical grain rotating at
angular velocity $\omega \ll \nu$
removes, on average, an angular momentum $\hbar\omega/(2\pi\nu)$.
The grain is heated by absorption of photons from the background
starlight, with energy density $u_*$, which we assume to have
the spectrum estimated for interstellar starlight by
Mezger, Mathis, \& Panagia (1982) and Mathis, Mezger, \& Panagia (1983).
The infrared emission cross section is assumed to scale
with frequency as
\beq
Q_\nu=Q_0\left(\frac{\nu}{\nu_0}\right)^2~~~,
\label{q}
\eeq
as appropriate for graphite, silicate, and various other candidate
grain materials (Draine \& Lee 1984).
Since there is reason to believe that the smallest interstellar
grains are predominantly carbonaceous,\footnote{
	The $10\micron$ silicate feature does not appear in the
	emission spectrum of diffuse interstellar clouds
	(Onaka, et al. 1996),
	implying that the particles small enough to be heated
	to $T_d\gtsim150\K$ by single-photon heating do not
	have a significant silicate component.
	}
here we will assume optical
properties appropriate for small graphitic particles (though
the discussion in Appendix \ref{app:IRdamping} is fully general).

If the grain is large enough that its temperature may be approximated
as a constant $T_d$, 
then the rotational drag is characterized by
(see Appendix \ref{app:IR,c})\footnote{
	\label{fn:IRc}
	The IR drag torque depends on $Qa^2\propto a^3$.
	The factor $(a/a_x)^4$ enters in (\ref{eq:Firc}) and
	(\ref{eq:Firq})
	from the normalization (\ref{eq:fdef}).
	}
\beq
F_{\IR,c} =
{60.8\over a_{-7}} \left({u_*\over u_\ISRF}\right)^{2/3}
\left({20\cm^{-3}\over n_\H}\right)
\left({100\K\over T}\right)^{1/2}
\left({a\over a_x}\right)^{4} ~~~.
\label{eq:Firc}
\eeq

For small grains, the quantized heating by starlight photons
results in thermal ``spikes'' followed by intervals of cooling
(see, e.g., Draine \& Anderson 1985),
thereby altering the emission spectrum and
affecting the rotational damping, as noted by Rouan et al (1992).
As shown in Appendix \ref{app:IR,q}, if individual heating events may
be treated separately, for very small carbonaceous grains we obtain
(cf. eq. \ref{eq:firq})
\beq
F_{\IR,q} = 
4.49 a_{-7}^{1/2}
\left({u_*\over u_\ISRF}\right)
\left({20\cm^{-3}\over n_\H}\right)
T_2^{-1/2}
%\left({100\K\over T}\right)^{1/2}
\left({a\over a_x}\right)^4
\label{eq:Firq}~~~.
\eeq
We have $F_{\IR,q}=F_{\IR,c}$ for $a=a_{\IR}$,
where
\beq
a_\IR = 5.68\times10^{-7}
\left({u_*\over u_\ISRF}\right)^{-2/9}\cm ~~~.
\eeq
We use eq.(\ref{eq:Firc}) for $a>a_{\IR}$, and
eq.(\ref{eq:Firq}) for $a<a_{\IR}$, or
\beq
F_\IR = \min(F_{\IR,q},F_{\IR,c}) ~~~.
\label{eq:F_IR}
\eeq
In Figs. \ref{fig:F_MC} -- \ref{fig:F_WIM} we show $F_\IR$ as a function
of grain radius $a$.

\subsection{Electric Dipole Damping}

The spinning grain will radiate power 
as given by eq. (\ref{eq:Pspin}).
The associated rotational damping time 
depends on $\omega$.
It will prove convenient 
(see \S\ref{sec:rot_rate})
to define a characteristic damping time 
\begin{eqnarray}
\ted &\equiv& {3 I^2 c^3 \over 4\mu^2kT}
=
{16\pi^2\xi^2\over75}
{\rho^2a^{10}c^3\over\mu^2kT}\\
&=&7.13\times10^{10}\s
{\xi^2a_{-7}^{8}\over 
( (a_x/a)^2 \langle Z^2\rangle+3.8(\beta/0.4\D)^2a_{-7})T_2}
\label{eq:ted} ~~~,
\end{eqnarray}
where we have assumed the dipole moment to be given by eq.\ (\ref{eq:mu2}).
With $\ted$ defined by eq.(\ref{eq:ted}), we see that electric
dipole damping contributes
\beq
\left({1 \over \omega}{d\omega\over dt}\right)_{ed} = 
- {I\omega^2\over 3kT}{1\over \ted} ~~~.
\eeq
In Figures \ref{fig:F_MC} -- \ref{fig:F_WIM} we show 
$\tau_H/\ted$ as
a function of $a$.

\subsection{Relative Importance of Damping Mechanisms}

Although all the processes discussed above contribute to rotational damping, 
some of them are more important than others. 
In {\it diffuse molecular
clouds} (see Fig. \ref{fig:F_MC}) the plasma drag is dominant.
This
result is counterintuitive as the degree of ionization 
is small ($x=10^{-4}$). 
To understand this, recall that $F_p \propto a^{-5}T^{-2}$, so that
for small
grains and low temperatures plasma drag can
dominate the gaseous drag, even when the fractional ionization is low.
Dipole emission drag is of
only marginal importance
because of the relatively high gas
density in the molecular region.
 
In {\it cold neutral media} (see Fig. \ref{fig:F_CNM}) 
plasma drag $F_p$ dominates 
in the range $6\times 10^{-8}$~cm$\ltsim a\ltsim 1.5\times 10^{-7}$~cm. 
For smaller grains
damping via dipole emission is more important, while for larger grains
infrared emission damping dominates. 
The infrared damping $F_{IR}\propto a^{-1}$ for large grains 
(with steady temperatures)
and $F_{IR}\propto a^{1/2}$ for small ones (which undergo
temperature ``spikes'').

In {\it warm neutral media} (see Fig. \ref{fig:F_WNM}) 
and in {\it  warm ionized media} (see Fig. \ref{fig:F_WIM})
the dipole emission is dominant for
grains $\lesssim 1.5 \times 10^{-7}$~cm mostly as the consequence of 
less efficient coupling between grains and the surrounding gas.
For larger grains infrared damping dominates. 
The difference between the 
two media evidently stems from the fractional ionization:
in ionized media interactions with ions and plasma effects are more
important than collisions with neutrals.

\section{Rotational Excitation\label{sec:rotational_excitation}}

\subsection{Recoil from Thermal Collisions and Evaporation}

An initially stationary grain will have its rotational
kinetic energy increasing at an average rate
\beq
\label{eq:gdef}
{d\over dt}\left({1\over2}I\omega^2\right)
=
n_\H\left({8kT\over\pi m_\H}\right)^{1/2} {2\pi a_x^4 m_\H kT \over I} 
(G_n+G_i+G_p+G_\IR) ~~~,
\eeq
where the normalized excitation rate
$G_n$ is due to impacting neutrals,
$G_i$ is due to impacting ions,
$G_p$ is due to plasma drag,
and
$G_\IR$ is due to recoil from infrared emission.

Neutrals deposit angular momentum when they impact the grain,
and give the grain an additional kick when they subsequently evaporate.
These processes (see Appendix \ref{app:F_n}) result in a normalized
excitation rate
\begin{eqnarray}
G_n =
\sum_{Z_g} f(Z_g)
\sum_n {n_n\over 2n_\H}
\left({m_n\over m_\H}\right)^{1/2}
\Biggl\{
\!\!\!\!&&\!\!\!\!
\exp(-Z_g^2\epsilon_n^2) + 2Z_g^2\epsilon_n^2 + \nonumber\\
&&\!\!\!\!{T_{ev}\over T}
\left[ \exp(-Z_g^2\epsilon_n^2 T/T_{ev}) + 2Z_g^2\epsilon_n^2 T/T_{ev}\right]
\Biggr\}
\label{eq:g_n} ~~~.
\end{eqnarray}

\begin{figure}
\epsscale{1.00}
\plotone{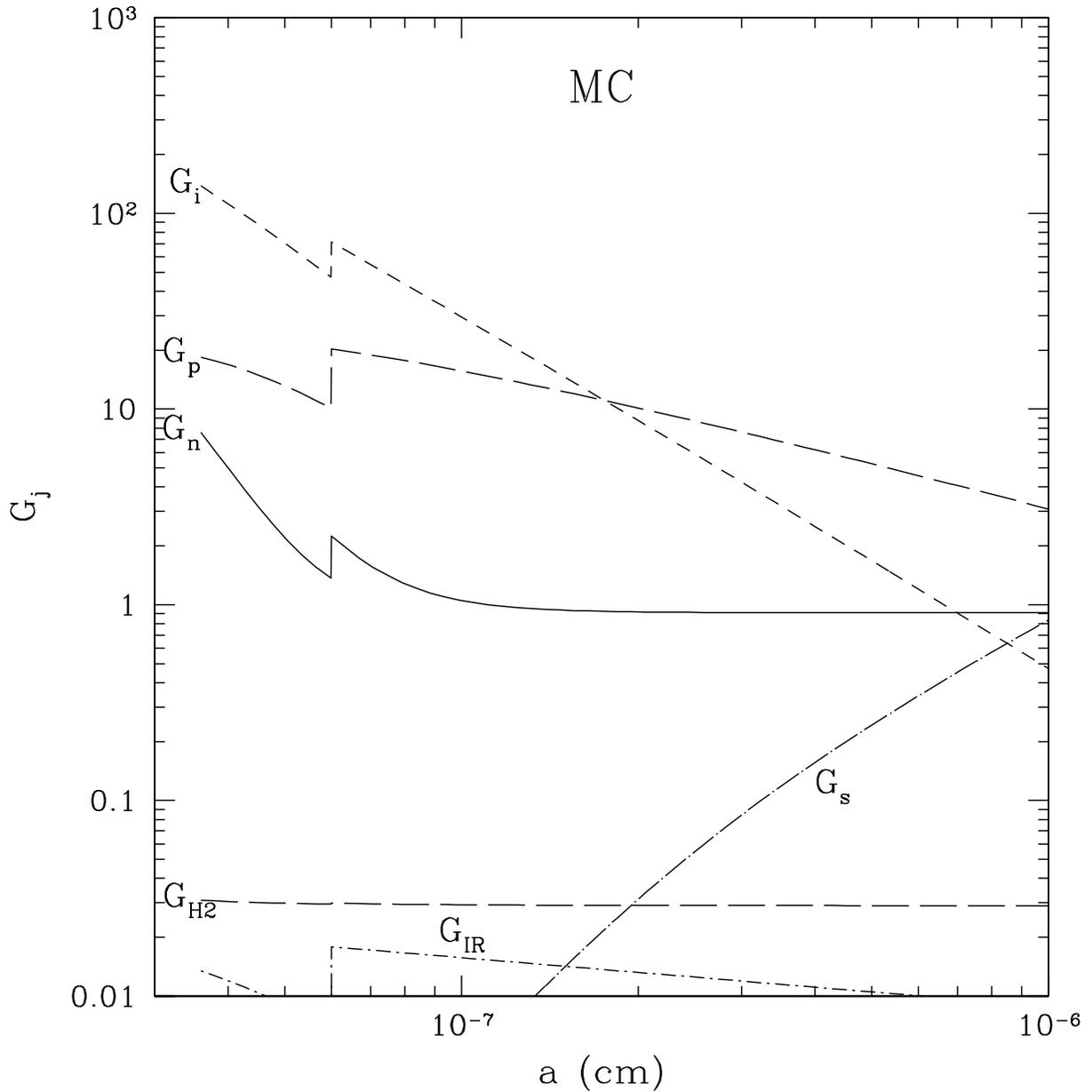}
\caption{
	\label{fig:G_MC}
	Dimensionless rotational excitation functions for neutral
	collisions [$G_n$, eq.(\protect{\ref{eq:g_n}})],
	ion collisions [$G_i$, 
	eq.(\protect{\ref{eq:g_i}})-(\protect{\ref{eq:giev}})],
	and plasma drag [$G_p$, eq.(\protect{\ref{eq:G_p}})]
	for CNM conditions.
	Ion collisions and plasma drag dominate, despite the
	low assumed ionization fraction of $10^{-4}$.
	Also shown is the ``superthermal'' excitation
	term $G_s$ for grains with recombination efficiency
	$\gamma=0.1$ and $N_r$ given by eq.\protect{\ref{eq:N_r}}.
	Because of the small assumed H fraction (1\%) $\HH$
	formation torques are not important in molecular clouds.
	}
\end{figure}

\begin{figure}
\epsscale{1.00}
\plotone{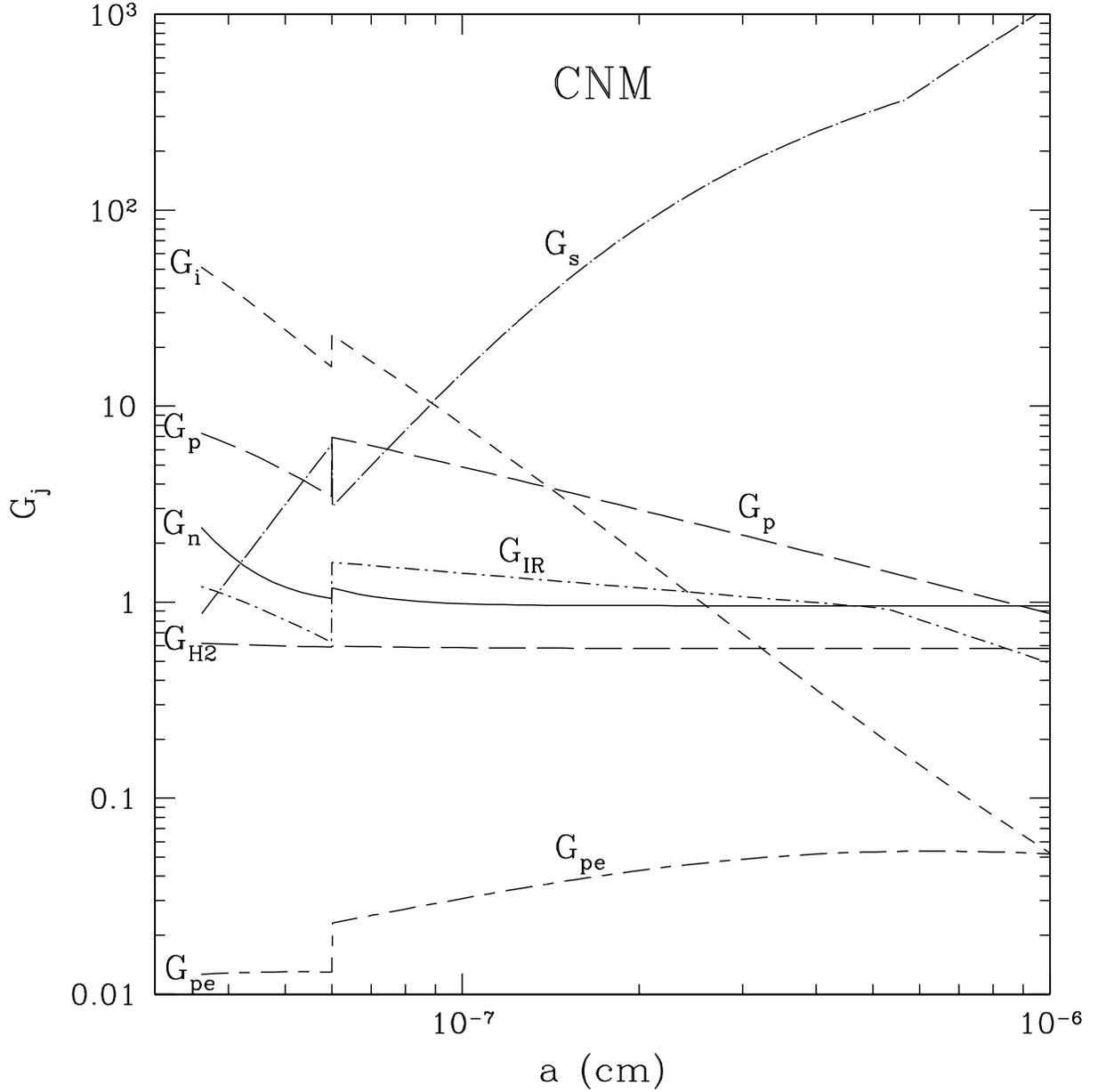}
\caption{
	\label{fig:G_CNM}
	Same as Fig. \protect{\ref{fig:G_MC}} but for
	CNM conditions.
	The contribution of infrared emission [$G_\IR$, 
	eq.(\protect{\ref{eq:g_irc}})-(\protect{\ref{eq:g_ir}})]
	is also shown.
	$G_{\rm H2}$ is the contribution if $\HH$ formation
	occurs randomly, rather than at preferred sites.
	$G_{pe}$ is the contribution from photoelectric 
	emission (eq. \protect{\ref{eq:G_pe}}).
	Ion collisions and plasma drag dominate for
	$a\ltsim10^{-7}\cm$, and systematic torques from
	$\HH$ formation dominate for larger grains for
	the assumed $\HH$ formation efficiency $\gamma=0.1$.
	}
\end{figure}

\begin{figure}
\epsscale{1.00}
\plotone{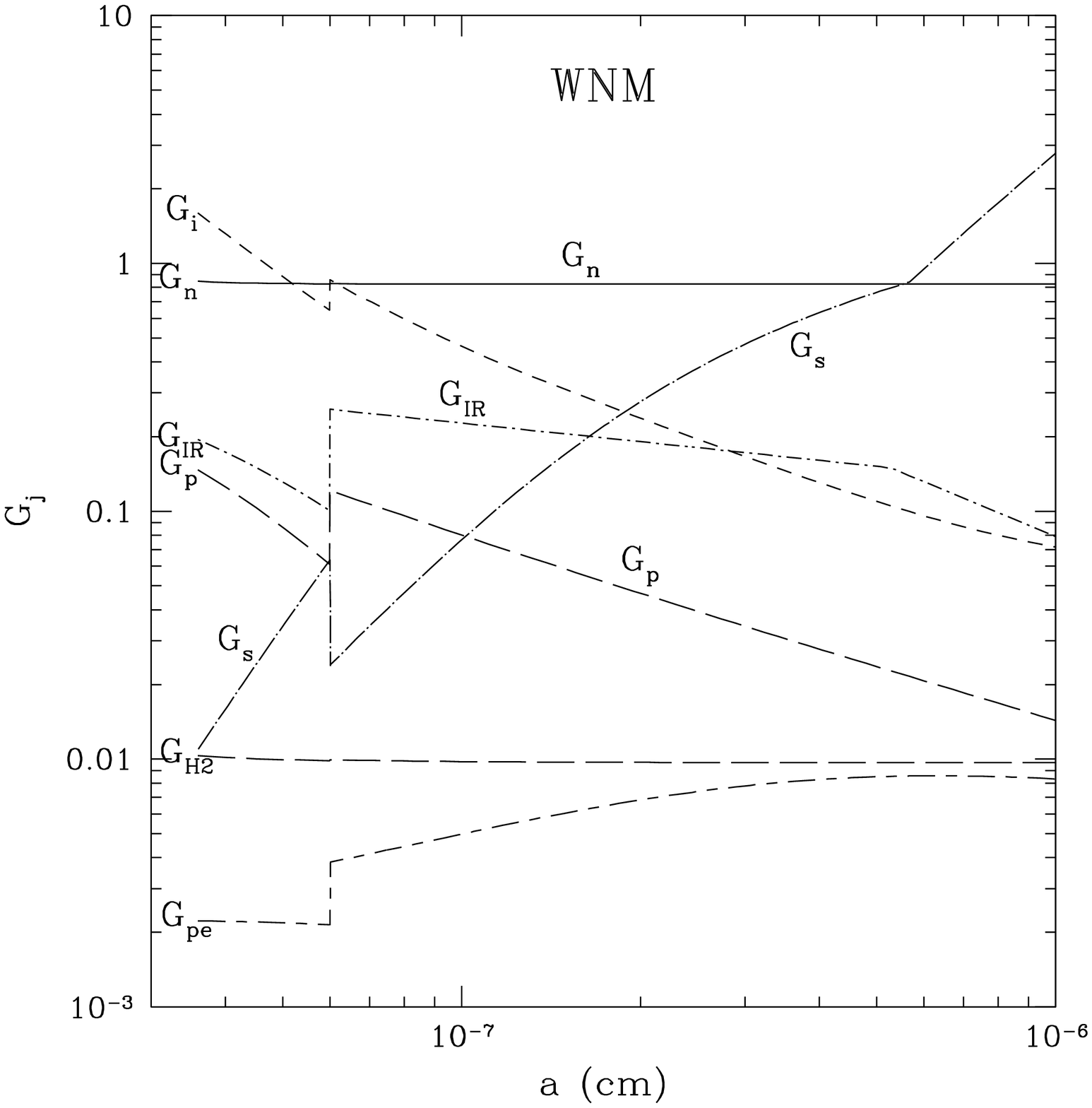}
\caption{
	\label{fig:G_WNM}
	Same as Fig. \protect{\ref{fig:G_MC}} and
	\protect{\ref{fig:G_CNM}} but for
	WNM conditions.
	}
\end{figure}

\begin{figure}
\epsscale{1.00}
\plotone{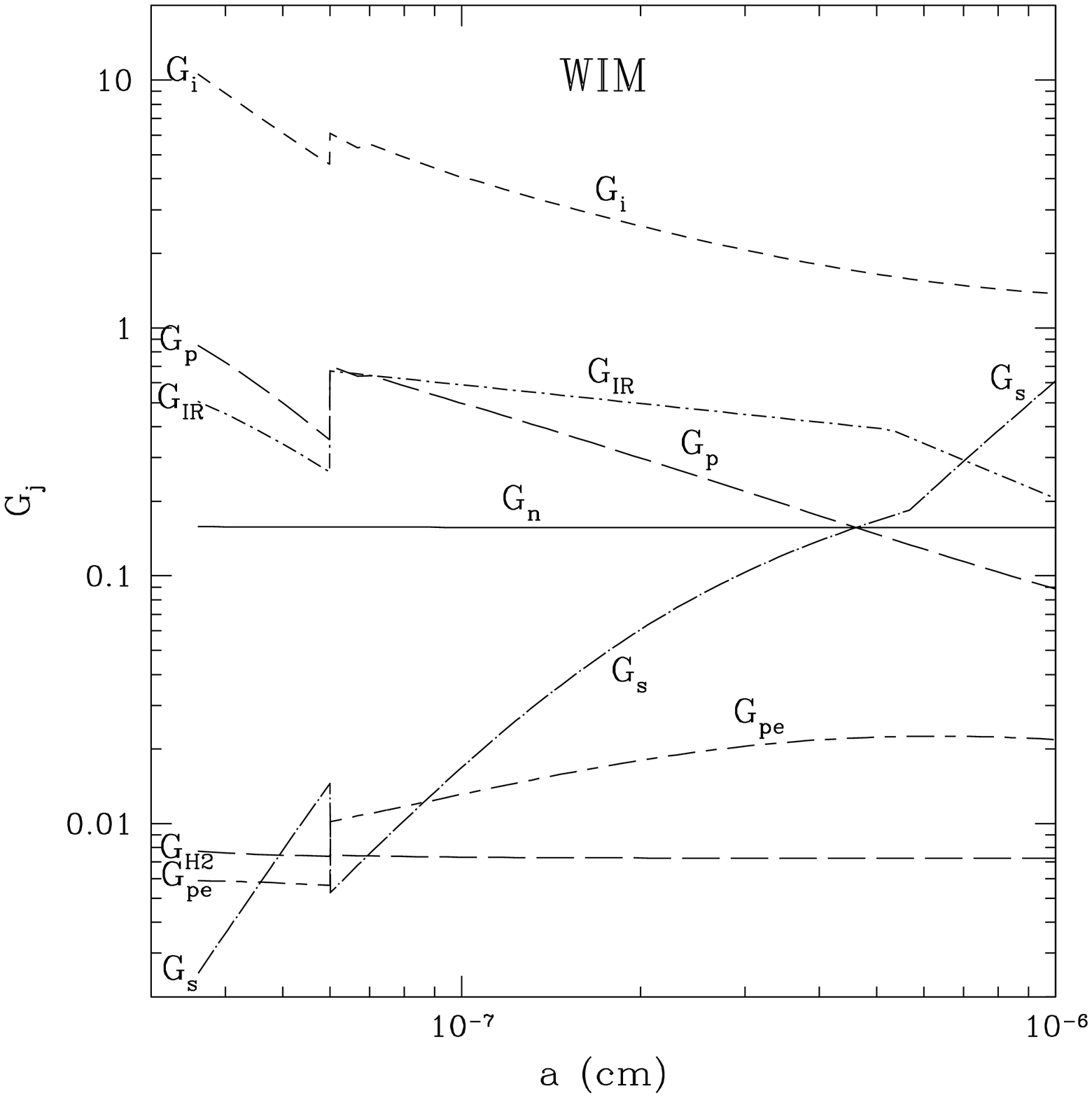}
\caption{
	\label{fig:G_WIM}
	Same as Fig. \protect{\ref{fig:G_MC}} and
	\protect{\ref{fig:G_CNM}} but for
        WNM conditions.
	}
\end{figure}

We separate the ion contribution into that delivered by incoming
ions, and that due to the evaporating neutrals resulting from 
the ion collisions:
\beq
G_i = G_i^{(in)} + G_i^{(ev)}
\label{eq:g_i}~~~,
\eeq
\begin{eqnarray}
G_i^{(in)} &=&
\sum_{Z_g\neq0} f(Z_g)
\sum_i {n_i\over 2n_\H}
\left({m_i\over m_\H}\right)^{1/2}
g_2 ( Z_iZ_ge^2/a_s kT) \nonumber\\
&&+ f(0)\sum_i {n_i\over 2n_\H}
\left({m_i\over m_\H}\right)^{1/2}
\left[
1+{3\pi^{1/2}\over4}\phi + {1\over2}\phi^2
\right]
\label{eq:giin}~~~,
\end{eqnarray}
\begin{eqnarray}
G_i^{(ev)} 
&=&
\sum_{Z_g\neq0} f(Z_g)
\sum_i {n_i\over 2n_\H}
\left({m_i\over m_\H}\right)^{1/2}
g_1 (Z_iZ_ge^2/a_s kT)
{T_{ev}\over T}\times\nonumber\\
&&~~~~~~~
\left[
{
\exp(-Z_g^2\epsilon_i^2) + 2Z_g^2\epsilon_i^2
\over
\exp(-Z_g^2\epsilon_i^2) +|Z_g|\epsilon_i \pi^{1/2} {\rm erf}(|Z_g|\epsilon_i)
}
\right]\nonumber\\
&&+ f(0)\sum_i {n_i\over 2n_\H}
\left({m_i\over m_\H}\right)^{1/2}
{T_{ev}\over T}\left[1+{\pi^{1/2}\over2}\phi\right]
\label{eq:giev}~~~,
\end{eqnarray}
\beq
g_2(x) = \left\{
\begin{array}{ll}
	1-x+x^2/2 & \mbox{if $x < 0$} \\
	e^{-x}	& \mbox{if $x >0$} 
\end{array}
	\right. ~~~,
\eeq
where we once again sum over neutrals $n$ and ions $i$.
$\phi$ is defined by eq.(\ref{eq:phidef}) and $\epsilon_i$ is defined
by eq.(\ref{eq:epsilondef}) but using the polarizability of the species
resulting from neutralization of the incoming ion $i$.
In Figs. \ref{fig:G_MC}-\ref{fig:G_WIM} we show $G_n$ and $G_i$
as functions of grain radius $a$.
Coulomb attraction makes $G_i$ the dominant excitation process
for small grain.

\subsection{Excitation by the Plasma}

Excitation of grain rotation by the fluctuating electric field from
passing ions is [cf. eq.(\ref{eq:F_p})]
\beq
G_p = {2e^2\mu^2\over 3a_x^4(kT)^2}
\sum_i {n_iZ_i^2\over n_\H}\left({m_i\over m_\H}\right)^{1/2}
\left\{
\ln(b_\omega/a)+\cos^2\Psi\ln\left[\min(b_q,\lambda_{\rm D})/b_\omega)\right]
\right\}
\label{eq:G_p}~~~.
\eeq

\subsection{Infrared Emission\label{sec:G_ir}}

Each infrared photon carries away angular momentum $\hbar$, and hence
there must be a corresponding change in the angular momentum of the
grain.
The energy in the radiated photons is due to absorption of starlight
photons, but since there are $\sim500$ infrared photons emitted per
starlight photon, we may neglect the angular momentum change due to the
photon absorption event itself.

If the grain temperature is approximated as constant and we assume
graphitic grains, then
(see Appendix \ref{app:IR,c})\footnote{
	The factor $(a/a_x)^4$ in eq.(\ref{eq:g_irc}) and (\ref{eq:g_irq})
	enters from the normalization of $G$
	[see eq.(\ref{eq:gdef})].}
\beq
G_{\IR,c}=
{7.34 \over a_{-7}}
\left({a\over a_x}\right)^4
\left({u_*\over u_\ISRF}\right)^{5/6}
\left({20\cm^{-3}\over n_\H}\right)
T_2^{-3/2}
\label{eq:g_irc}~~~.
\eeq

For very small grains, the grain temperature may be treated as zero except
immediately following a photon absorption event, in which case for
graphitic grains we find:
\beq
G_{\IR,q}={2.11\over a_{-7}^{1/4}}
\left({a\over a_x}\right)^4
\left({u_*\over u_\ISRF}\right)
\left({20\cm^{-3}\over n_\H}\right)
T_2^{-3/2}
\label{eq:g_irq}~~~.
\eeq
We take
\beq
G_\IR = \min( G_{\IR,c},G_{\IR,q} ) ~~~.
\label{eq:g_ir}
\eeq

\subsection{Photoelectric Emission \label{sec:peemission}}

The contribution to rotational excitation by photoelectrons
emitted randomly from the grain is
\beq
G_{pe}={m_e\over 4n_\H(8\pi m_\H kT)^{1/2}a_x^2 kT}
\sum_{Z_g} f(Z_g)\dot{N}_{pe}
\left(\langle E_{pe}\rangle + {(Z_g+1)e^2\over a_s}\right)~~~,
\label{eq:G_pe}
\eeq
where $\dot{N}_{pe}(Z_g)$ is the rate of emission of escaping
photoelectrons from a grain with charge $Z_ge$, and
$\langle E_{pe}\rangle$ is the mean kinetic energy of the
escaping photoelectrons at infinity.
We assume the photoelectrons emerge from the grain surface
with a ``cosine-law'' directional distribution.
$\langle E_{pe}\rangle$ is as estimated by Bakes \& Tielens (1994).
Because $m_e\ll m_\H$, the contribution of photoelectric
emission to rotational drag is negligible.

\subsection{Random H$_2$ Formation \label{sec:randomH2}}

If a fraction $\gamma$ of arriving H atoms recombine at random on
the grain surface, and the nascent $\HH$ has an average
translational kinetic energy $E_{\rm f}$, then
the contribution to rotational excitation is
\beq
G_{\HH} = {\gamma\over4}(1-y){E_{\rm f}\over kT}
\left( 1+ 
{\langle J(J+1)\rangle \hbar^2 \over 2 m_\H E_{\rm f}a_x^2}
\right)~~~.
\label{eq:G_H2}
\eeq
The nascent $\HH$ is assumed to leave the grain surface with
a ``cosine-law'' directional distribution, with the
internal angular momentum $J\hbar$ uncorrelated with the velocity.

The efficiency $\gamma$ of H$_2$ formation on very small grains is very
uncertain, as is the effective number $N_r$ of active recombination
sites.
For very small grains, $\HH$ formation may be suppressed due to
temperature fluctuations (which limit the residence time
of H atoms) or chemical ``poisoning'' (Lazarian 1995) of recombination
sites, but the details are very uncertain.
The empirical rate coefficient for $\HH$ formation on grains
$R\approx3\times10^{-17}n(\H)n_\H\cm^3\s^{-1}$ (Jura 1975) provides a limit:
if small grains contribute a geometric cross section per H
$\sim2\times10^{-21}\cm^2/\H$
(see \S\ref{sec:sized}), then $\gamma\ltsim0.2$ if
$\HH$ formation on small grains is not to exceed the empirically-determined
total $\HH$ formation rate in the CNM.

In Figs. \ref{fig:G_MC}-\ref{fig:G_WIM} we show $G_{\HH}$ for
$\gamma=0.1$, $E_{\rm f}=0.2\eV$, and $\langle J(J+1)\rangle=10^2$.
We see that $G_{\HH}$ does not make a major contribution to
rotational excitation.

\subsection{Systematic Torques\label{sec:systematic}}

Superthermal rotation may result if systematic torques act on a 
grain. Purcell first showed that various processes,
including variations of the accomodation coefficient, photo-electric yield,
and formation of hydrogen on preferential sites, can drive suprathermal
rotation. Recently, radiative torques were identified as an important
means of suprathermal spin-up (Draine \& Weingartner 1996).

Thus far, suprathermal rotation has been discussed mostly
in relation to grain alignment and for relatively large grains,
i.e. with $a>10^{-6}$~cm (see Lazarian \& Draine 1997).
The degree of suprathermality, i.e. the ratio of grain kinetic energy to
its energy of thermal rotation, depends on grain size. 
%For instance,
%for the typical interstellar radiation field, torques due to starlight are 
%negligible for grains with $a \ll 10^{-5}$~cm. 

Systematic torques differ from the random torques in having a nonzero
time average (in grain coordinates).
Let $\Gamma_s$ be a time-averaged systematic torque in grain coordinates.
If the grain were not subject to fluctuating torques, it would attain
a steady rotational velocity $\omega_s$ satisfying
\beq
\Gamma_s = {I F \omega_s \over \tH} + {4\mu^2\omega_s^3\over 9c^3}~~~.
\eeq
For our purposes it is sufficient to use the simple approximation
\beq
\omega_s = {\Gamma_s \tH \over I F}
\left[
1 + {4\mu^2\over 9\Gamma_s c^3}\left({\Gamma_s\tH\over I F}\right)^3
\right]^{-1/3}
\label{eq:omega_s}~~~,
\eeq
which is exact in the limits $\Gamma\rightarrow0$ and
$\Gamma_s\rightarrow\infty$.
The rate at which the systematic torque does work on the grain is
then just $\Gamma_s\omega_s$.

The actual grain rotational velocity $\omega$ is also the result of
fluctuating torques, but $\Gamma_s\omega_s$ remains a good estimate
for the mean rate at which the systematic torque contributes to the
rotational kinetic energy.
Thus it is natural to 
characterize the importance of superthermal torques through
a dimensionless quantity $G_s$ defined by analogy with 
the dimensionless excitation rates $G_n$, $G_i$,
and $G_p$:
\beq
G_s \equiv {I\Gamma_s\omega_s \over n_\H (2\pi m_\H kT)^{1/2} 4 a_x^4 kT}
\label{eq:G_s}~~~,
\eeq
where $\omega_s$ is given by eq. (\ref{eq:omega_s}).

Formation of $\H_2$ molecules on catalytic sites can drive superthermal
rotation.
Averaged over grains of radius $a$, the rms torque is\footnote{
	The rms moment-arm $\propto (a_x^4/a_s^2)^{1/2}$, and the
	H$_2$ formation rate per site $\propto a_s^2/N_r$, hence
	$\Gamma_s\propto a_x^2a_s$.
	}
\beq
\Gamma_s = {4\gamma n(\H) a_x^2 a_s \over N_r^{1/2}} 
\left(\pi m_\H E_{\rm f} kT\right)^{1/2} ~~~,
\eeq
where $\gamma$ 
is the fraction of impinging H atoms which leave the grain
surface as H$_2$, 
$E_{\rm f}=0.2\hat{E}_{\rm f}\eV$ is the translational kinetic energy per
nascent H$_2$, 
$N_r$ is the number of recombination sites on the
grain surface, 
and we have assumed that at each recombination site
the newly-formed H$_2$ leaves the surface in a fixed direction, with
the direction for each site 
randomly-drawn from a ``cosine-law'' distribution.
Then, using eq.(\ref{eq:omega_s},\ref{eq:G_s})
\beq
G_s
\approx
1.75\times10^4 
{a_{-7}^3\over N_r} 
\left({a_s\over a}\right)^2\left({a\over a_x}\right)^4
{ \xi\hat{E}_{\rm f}\gamma^2 \over F T_2}
(1-y)^2
%\left({n(\H)\over n_\H}\right)^2
\left[
1 + {4\mu^2\over 9\Gamma_s c^3}\left({\Gamma_s\tH\over I F}\right)^3
\right]^{-1/3}
\label{eq:G_sH2}~~~.
\eeq

For large grains, we expect $N_r\propto a^2$,
but for all grains where $\HH$ forms ($\gamma >0$) we must
have $N_r\geq1$.
We will consider two extreme cases:
\begin{enumerate}
\item No H$_2$ formation: $\gamma=0$, $\Gamma_s=0$, and $G_s=0$.
\item H$_2$ formation with $\gamma=0.1$, $\hat{E}_{\rm f}=1$, and
the number of recombination sites on a grain taken to be
\beq
N_r=1+\left({a\over2\times10^{-7}\cm}\right)^2 ~~~.
\label{eq:N_r}
\eeq
\end{enumerate}

%Variations of the accommodation coefficient can be shown to be of marginal
%importance for small grains with $a<10^{-7}$ cm.
%Purcell (1979) showed that the resulting superthermality 
%is
%{\bf [CHECK THIS]}
%\beq
%\Delta G_s
%\approx 13 \left(\frac{s\delta}{a}\right)^2 a^3_{-7}
%\left({T-T_d\over T}\right)^2 F^{-2}
%\label{eq:G_accom}~~~,
%\eeq
%where 
%$s$ is the size of a patch with the variation of the accommodation
%coefficient $\delta$.
%For $s\delta/a<0.1$ and $a_{-7}\ltsim1$, $\Delta G_s \ll 1$ and
%can be neglected.
%
%Variations of the photo-electric emissivity over the grain surface
%will also provide a systematic torque, of the same order as the 
%one arising from
%variations of the accommodation coefficient
%(see Purcell 1979).

\subsection{Relative Importance of Excitation Mechanisms}

Unlike the case of the damping mechanisms, ion bombardment dominates
rotational excitation for small grains in all ISM phases discussed above.
This apparent disparity between damping and excitation processes,
which seemingly contradicts the Fluctuation-Dissipation theorem,
is the consequence of the non-equilibrium state of the
interstellar medium.
The ionization in $T\ltsim 10^4\K$ regions is primarily due to
photoionization and cosmic rays, and far exceeds the ionization which
would correspond to thermodynamic equilibrium at the kinetic temperature
of the gas.
The dominance of ion excitations stems from the increase
of both capture cross section and angular momentum per capture.

In {\it molecular clouds}  (see Fig. \ref{fig:G_MC})
excitation through ion collisions is dominant
for grains $\lesssim 1.7\times 10^{-7}$~cm.
As a result, these
grains rotate with velocities a bit {\it larger} than would
result from Brownian motion.

In {\it cold neutral media} the picture is more involved 
(see Fig.\ \ref{fig:G_CNM}). 
Regular torques associated with H$_2$ formation can affect grain dynamics
for $a\gtrsim 10^{-7}$~cm. The rate of excitation through infrared
emission is very close to that  of gaseous bombardment in the range
$4\times 10^{-8}$~cm$<a<4\times 10^{-7}$~cm and only for larger grains do
grain-neutral interactions dominate grain excitation.

The relative importance of ion excitations drops with grain temperature.
Therefore for {\it warm neutral media} (see Fig.\ \ref{fig:G_WNM}) 
collisions with ions constitute
the leading excitation mechanism only for grains less than $\sim
10^{-6}$~cm. For larger grains collisions with neutrals dominate.
However, for {\it warm ionized media} (see Fig.\ \ref{fig:G_WIM}), 
where 99\% of species
are ionized, collisions with ions are bound to be the major source
of rotational excitation.

\section{Rate of Rotation\label{sec:rot_rate}}

With $F$ defined by equation (\ref{eq:fdef}),
the time-averaged rate of change of rotational kinetic energy may be written
\beq
0 = \left\langle {d\over dt} {I\omega^2\over2}\right\rangle 
= {3kT\over \tH} G
- I{\langle\omega^2\rangle\over \tH} F
-{4\mu^2\over 9c^3}\langle\omega^4\rangle
\label{eq:rotrate}~~~,
\eeq
where
\beq
G \equiv G_n + G_i + G_p + G_\IR + G_s ~~~,
\label{eq:gsum}
\eeq
where $G_s>0$ if there are systematic torques due to 
$\H_2$ formation [eq.(\ref{eq:G_sH2})].
Although the distribution of angular velocities 
will not be Maxwellian (except in the limit where electric dipole emission
and superthermal torques are both negligible),
we assume the relation between moments of a Maxwellian distribution,
\beq
\langle\omega^4\rangle = {5\over3}\langle\omega^2\rangle^2 ~~~,
\eeq
to obtain a quadratic equation in $\langle\omega^2\rangle$ from
eq.\ (\ref{eq:rotrate}).
With the electric dipole damping time $\ted$ defined by eq.\ (\ref{eq:ted}),
the solution is
\beq
\langle\omega^2\rangle
=
{2 \over 1 + [1+ (G/F^2)(20\tH/3\ted)]^{1/2}}
\left({G \over F}\right)
\left({3kT\over I}\right)
\label{eq:<omega^2>}~~~.
\eeq
Eq. (\ref{eq:<omega^2>}) 
provides a reasonable estimate for the rms grain rotation rate.

The effective rotational damping time $\tau_J$ is given by
\beq
{1\over\tau_J} = {F \over \tH} + 
\left( {I\omega^2\over 3kT}\right) {1\over\ted}
\label{eq:tauJ}~~~.
\eeq

%In the present application we are concerned with the effective rotation
%rate $\omega_{rad}$ for electric dipole radiation from grains of radius $a$:
%\beq
%\omega_{rad} = \langle \omega^4 \rangle ^{1/4} \approx
%\left({5 \over 3}\right)^{1/4}
%\left\{{2 \over 1 + [1+ (G/F^2)(20\tH/3\ted)]^{1/2}}
%\left({G \over F}\right)
%\left({3kT\over I}\right)
%\right\}^{1/2}~~~.
%\eeq

\section{Effects of Impulsive Torques\label{sec:impulsive_J}}

The above discussion of grain rotation has been directed at estimating
$\langle\omega^4\rangle$ for grains of radius $a$.
%and it has been assumed
%that the emissivity of grains of radius $a$ is a $\delta$-function at
%$\langle\omega^4\rangle^{1/4}=
%(5/3)^{1/4}\langle\omega^2\rangle^{1/2}=
%1.136\langle\omega^2\rangle^{1/2}$.
%This approximation is clearly inexact, but if the distribution of
%angular velocities is Maxwellian with temperature $T_r$, the emissivity
%$dP/d\omega \propto \omega^6 \exp(-I\omega^2/2kT_r)$ will peak
%at 
%$\omega=(6kT_r/I)^{1/2}=1.414\langle\omega^2\rangle^{1/2}$.
%
Because we do not have thermal equilibrium, however, the angular velocity
distribution may depart substantially from a Maxwellian.
In particular, if $I\delta \omega$ 
is the angular momentum of an impacting particle,
then if $\langle\delta \omega^2\rangle \gtsim \langle\omega^2\rangle$,
then the angular velocity distribution will depart from a Maxwellian.
This occurs when infrared emission and/or electric
dipole emission is able to reduce the grain angular velocity substantially
between impact events.

\begin{figure}
\epsscale{1.00}
\plotone{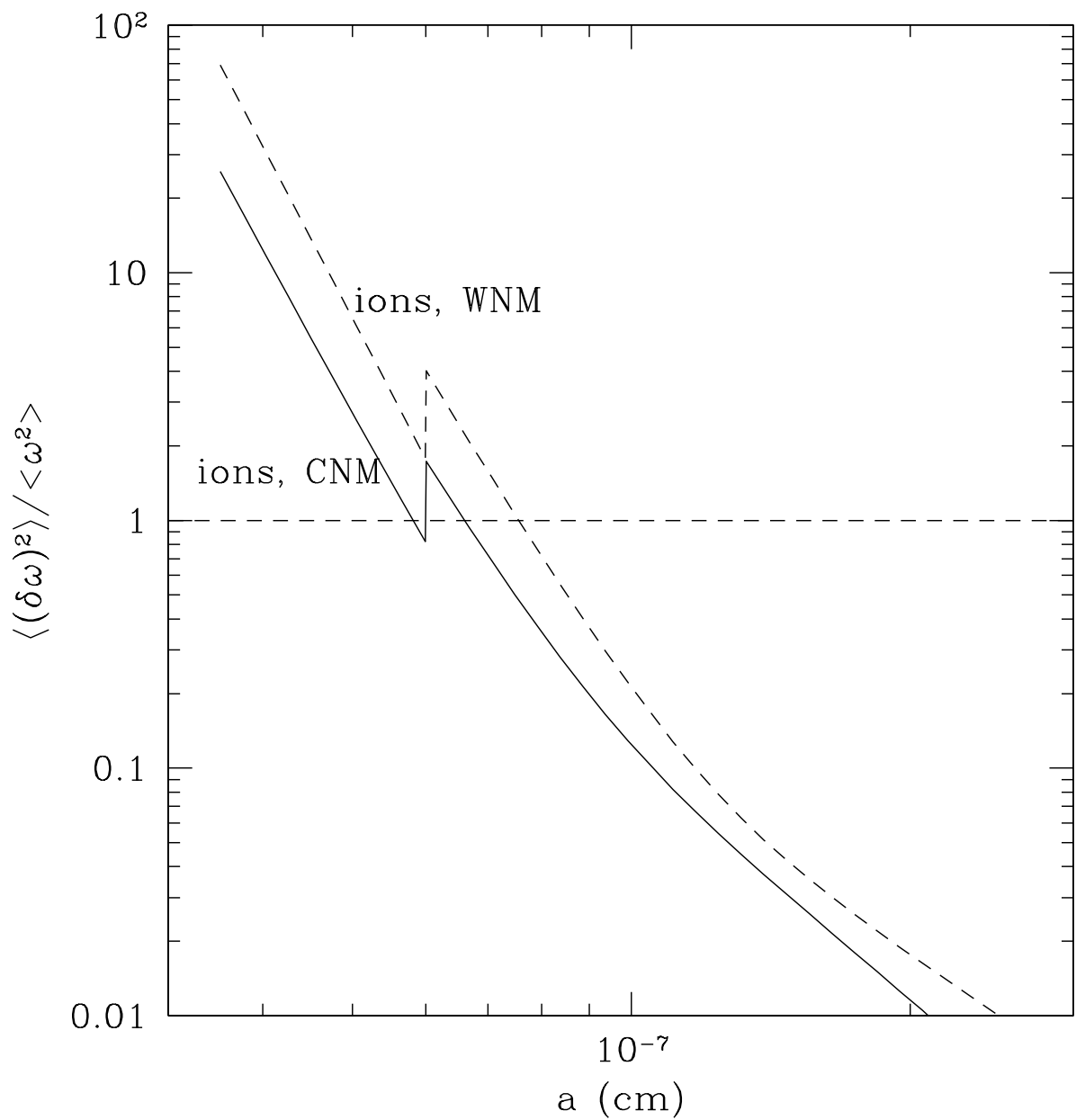}
\caption{
	\label{fig:dJ2_over_J2}
	Mean square $\delta \omega$ due to impacting ions 
	relative to mean square $\omega$ as a function
	of grain radius $a$, under CNM and WNM conditions.
	It is seen that the rotational excitation process may be treated
	as continous for $a\gtsim 7\Angstrom$, but for smaller grains a
	single ion impact can result in an angular momentum much larger
	than the time average.
	For these small grains the angular velocity distribution will
	depart significantly from a Maxwellian.
	}
\end{figure}

One can show that colliding species $j$ has
\beq
\langle(\delta \omega)^2\rangle_j = (G_j/F_j) 2a^2m_j kT/I^2 ~~~.
\eeq
In Figure \ref{fig:dJ2_over_J2} we show $\langle(\delta \omega)^2\rangle$ for
impacting ions, normalized by $\langle\omega^2\rangle$ for the grain.
We see that for large grains, the impulsive nature of the rotational excitation
is unimportant, and the time evolution of the grain rotation can be
treated as a continuous process.
For grains with $a\ltsim 7\Angstrom$, however, a single ion impact can change
the grain angular momentum substantially.
From Fig. \ref{fig:G_CNM} we see that ion impacts are the dominant source
of rotational excitation for small grains in the MC, CNM, and WIM phases.
This is primarily the result of the strong electric dipole damping for small
grains (see Figs. \ref{fig:F_CNM} -- \ref{fig:F_WIM}) in diffuse gas, which 
causes the smaller grains to have sub-thermal rates of rotation.

\section{Centrifugal Stresses\label{sec:centrifugal}}

\begin{figure}
\epsscale{1.00}
\plotone{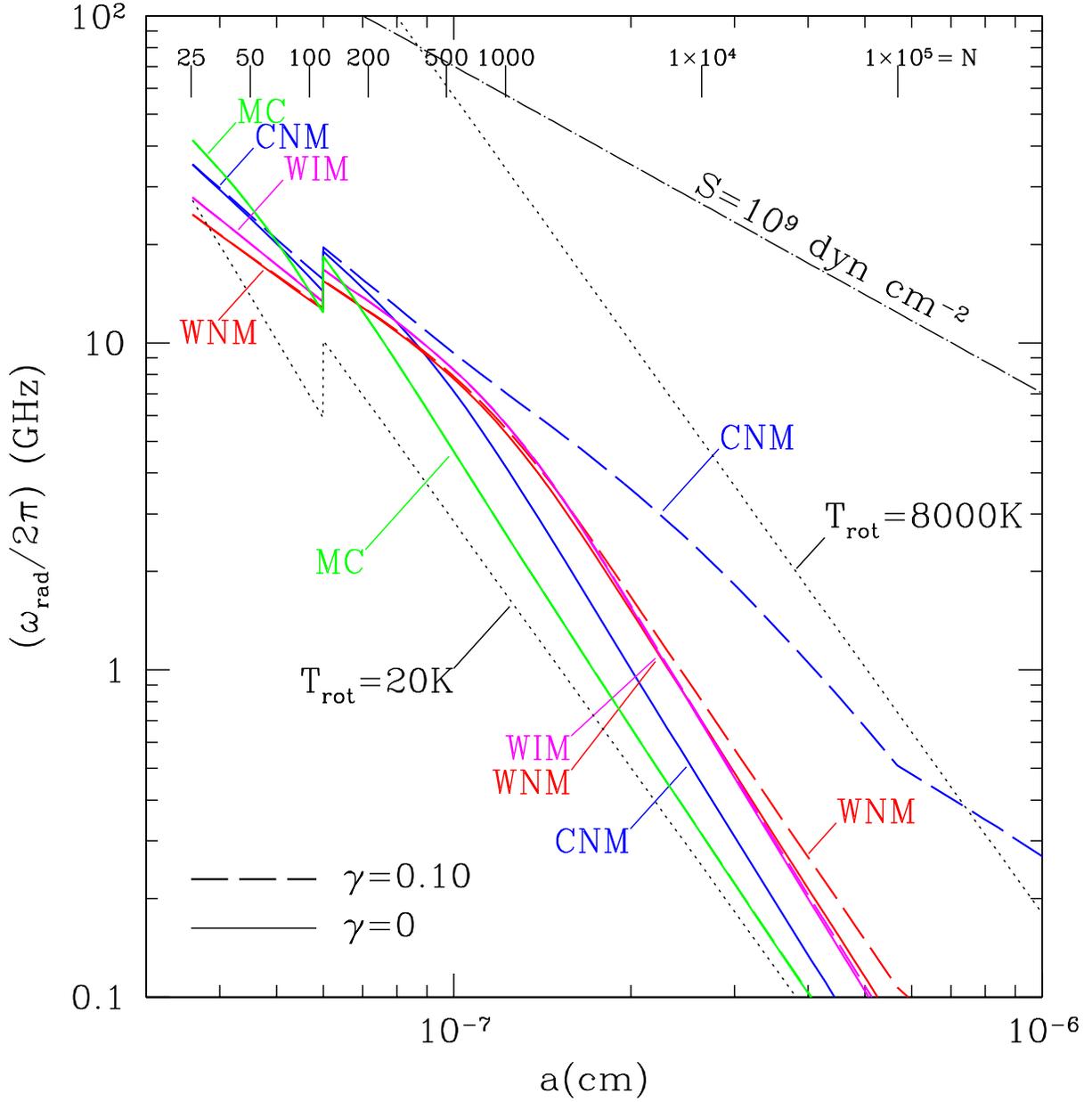}
\caption{
	\label{fig:omega}
	Effective rotation rate $\omega_{rad}=\langle\omega^4\rangle^{1/4}$ 
	as a function of
	grain radius $a$, for various environmental conditions,
	and with and without $\HH$ formation.
	Also shown are the thermal rotation rates
	[eq. (\protect{\ref{eq:nurms}})]
	at $T=20\K$ and $8000\K$.
	The number of atoms $N$ in a grain is
	indicated at the top of the figure.
	Also shown is the rotation rate for which the
	tensile stress equals $10^9{\rm ~dyn}\cm^{-2}$.
	}
\end{figure}

An upper limit for the grain rotational frequency $\omega$ can be obtained
from the ability of grains to withstand centrifugal stress
(Draine \& Salpeter 1979).
With a characteristic stress $(1/4)\rho\omega^2 a_x^2$, and a maximum
stress $S_{max}$, the maximum rotational frequency is
\beq
{\omega\over2\pi}\ltsim {1\over\pi a_x} \left({S_{max}\over\rho}\right)^{1/2}
= 7\times10^{10}{1\over a_{-7}}\left({a\over a_x}\right)
\left({S_{max}\over 10^9\erg\cm^{-3}}\right)^{1/2}\Hz~~~.
\eeq
Bulk
polycrystalline substances have
$S_{max}\approx 10^{9}\erg\cm^{-3}$, while 
for single-crystal materials 
$S_{max}\approx 10^{11}\erg\cm^{-3}$.
As can be seen from Figure \ref{fig:omega}, 
the expected rotation rates are such that even grains
with $S_{max}$ as small as $10^9\erg\cm^{-3}$ can survive.

\section{Size Distribution\label{sec:sized}}

\subsection{Diffuse Clouds}

As shown below, microwave emission from grains is only significant
for $\omega/2\pi\gtsim 1 \GHz$, so from Fig. \ref{fig:omega} we see
that only $a\ltsim3\times10^{-7}\cm$ grains ($N\ltsim10^4$) are of
capable of contributing to such emission.

The size distribution of these ultrasmall dust grains is very poorly known.
Studies of interstellar extinction are relatively insensitive to the
detailed distribution of very small dust grains.
Observations of extinction at $\lambda\gtsim 1000\Angstrom$
constrain the total mass in dust at $a\ltsim 1\times10^{-6}\cm$, but
do not reveal how this mass is distributed over grains of different
sizes.

The MRN distribution (Mathis, Rumpl, \& Nordsieck 1977),
extended down to very small sizes, provides one
estimate for the population of very small grains.
The graphitic component has
\beq
{dn\over da} = n_\H A_\MRN a^{-3.5}~~~
\mbox{for $a_{min}<a<2.5\times10^{-5}\cm$},
\label{eq:sizedmrn}
\eeq
with $A_\MRN=10^{-25.16}\cm^{2.5}$ (Draine \& Lee 1984).
This size distribution,
if extended down to $a_{min}=3.6\Angstrom$,
has 17\% of the grain mass in $a<10^{-6}\cm$ grains,
but only 2.6\% of the mass is in $a<10^{-7}\cm$ grains.
This distribution of graphitic grains would
contribute a geometric cross section per H atom
$\Sigma_\MRN=2.2\times10^{-21}\cm^2$.

Observations of the ``UIR'' emission features, as well as the strong
$12\micron$ and $25\micron$ emission observed by IRAS 
(Boulanger \& P\'erault 1988),
have been interpreted as showing that there must be a substantial
population of very small grains 
(Leger \& Puget 1984; Draine \& Anderson 1985).
Here we consider a log-normal size distribution for this population, which
we add to the MRN distribution:
\beq
{1\over n_\H}{dn\over da} = A_{MRN} a^{-3.5} + 
B a^{-1}\exp\left[-{1\over2}\left({\ln(a/a_0)\over\sigma}\right)^2\right]
\label{eq:sized}~~~.
\eeq
We will consider $a_0=6\Angstrom$ and $\sigma=0.4$;
The coefficient $B$ is chosen so that
the grains contain a fraction $f_\C=0.05$ of the total carbon
abundance ($\C/\H=4.0\times10^{-4}$: Grevesse et al. 1991).
This grain component then contributes an additional geometric
cross section per H atom $\Sigma=1.7\times10^{-21}\cm^2$.

For comparison, we note that the grain model of D\'esert, Boulanger,
and Puget (1990) has $f_\C=0.09$ in the ``PAH'' component, molecules with
$60<N<540$ C atoms.

\subsection{Dense Clouds}

The abundance of ultrasmall grains in dense interstellar clouds is not
well known.
In denser regions, the wavelength $\lambda_{max}$
of maximum polarization is increased, and the ratio 
$R_V\equiv A_V/E(B-V)\approx5.5$, 
significantly larger than the value $R_V\approx3.1$ characteristic
of diffuse regions; both are indications that the fraction of the grain
mass in smaller grains is reduced in denser regions.
Here we will assume that the fraction of the mass in ultrasmall
grains is reduced by a factor 5 relative to diffuse clouds:
for $a\ltsim3\times10^{-7}\cm$ 
we adopt the size distribution (\ref{eq:sized}) but with $A_{MRN}$
and $B$ reduced by a factor 5 from the values used for diffuse clouds.

\section{Emissivity \label{sec:emissivity}}

Recognizing that there will be a range of dipole moments among grains
of a given size,
we will assume that
25\% of the grains have $\beta=0.5\beta_0$, 50\% have $\beta=\beta_0$,
and 25\% have $\beta=2\beta_0$, with $\beta_0=0.4\debye$ as our
standard value.

\begin{figure}
\epsscale{1.00}
\plotone{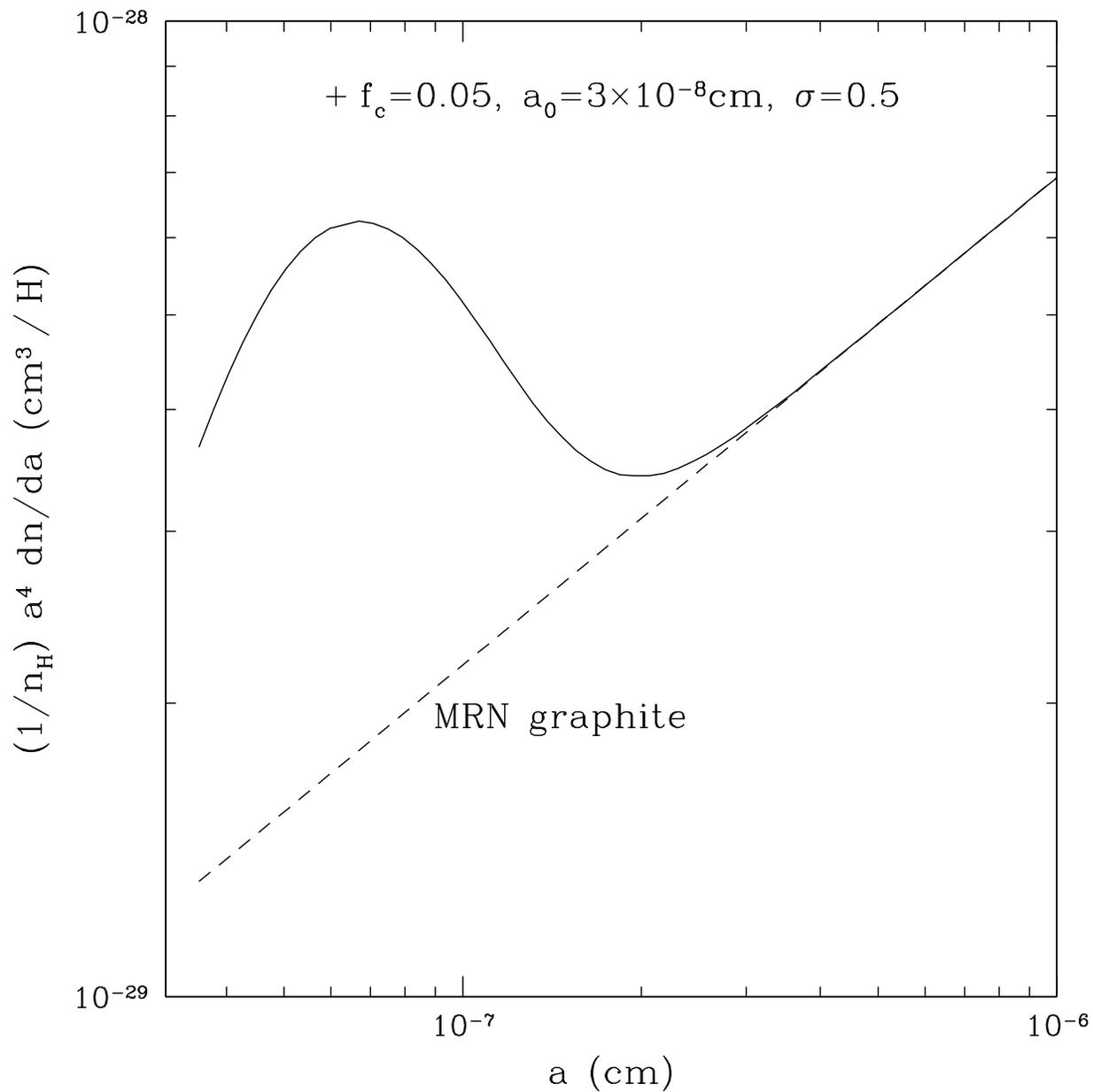}
\caption{
	\label{fig:sized}
	Adopted sized distribution for small grains in diffuse
	regions (CNM,WNM,WIM), with 5\% of the
	total carbon abundance in a log-normal component
	(see eq. \protect{\ref{eq:sized}}).
	In molecular gas, we assume the number of $a\ltsim3\times10^{-7}\cm$
	grains to be reduced by a factor 5 (see text), so that the
	log-normal component contains 1\% of the carbon.
	}
\end{figure}

For each grain size $a$ and value of $\beta$,
there is 
%a mean square dipole moment $\mu^2$ and
a mean square angular velocity $\langle\omega^2\rangle$.
If we assume $\omega$ to follow a Boltzmann distribution, then
the emissivity per H is
\beq
\label{eq:emissivity}
{j_\nu\over n_\H} = 
\left({8\over 3\pi}\right)^{1/2}{1\over n_\H c^3}
\int da {dn\over da}{\mu^2\omega^6\over \langle\omega^2\rangle^{3/2}}
\exp\left({-3\omega^2\over2\langle\omega^2\rangle}\right) ~~~.
\eeq
We evaluate this emissivity for the size distribution of Fig. \ref{fig:sized},
for conditions characteristic of different phases of the
interstellar medium.
We consider the case where no $\HH$ formation takes place, as well as
the case where it occurs with 10\% efficiency ($\gamma=0.1$).

Ferrara \& Dettmar (1994) predicted the microwave emission from grains
under WIM conditions, obtaining an emissivity 
$j_\nu/n_\H=1.1\times10^{-15}(\nu/100\GHz)^{2.8}\Jy\sr^{-1}$.
The slope and magnitude of this emission is similar to our results near
$10\GHz$, but this is coincidental: Ferrara \& Dettmar assumed grains
to be rotating with thermal rotation rates corresponding to
$\sim10^4\K$, whereas we find very strong damping by 
both microwave emission and infrared emission (see Fig.\ \ref{fig:F_WIM})
so that the grains actually rotate very subthermally (relative to the $8000\K$
gas temperature in the WIM).
For example, for $a=10^{-7}\cm$ grains under WIM conditions,
we estimate $\omega_{rad}/2\pi\approx 8.5 \GHz$
(see Fig.\ \ref{fig:omega}), whereas
Ferrara \& Dettmar estimate a rotation rate $\sim50\GHz$ for this size;
since the emitted power $\propto \omega^4$, this is a significant
difference.
In Ferrara \& Dettmar's model, the 10 GHz emission comes from grains
with $a\approx2\times10^{-7}\cm$, whereas we find here that
the 10 GHz emission comes primarily from grains with 
$a\approx9\times10^{-8}\cm$;
these two sizes differ by a factor of $\sim10$ in mass, and 
$\sim50$ in moment of inertia.

\begin{figure}
\epsscale{1.00}
\plotone{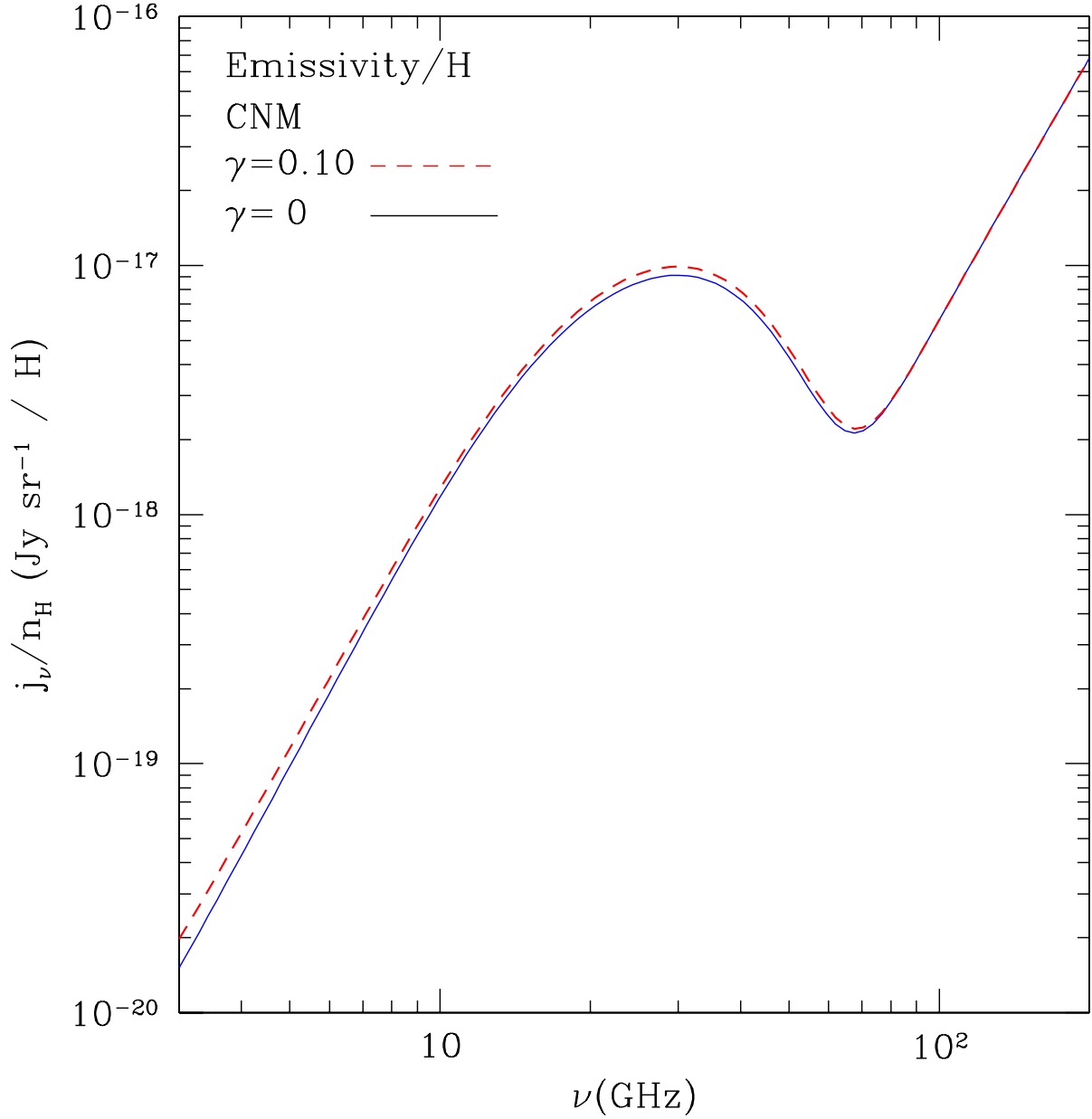}
\caption{
	\label{fig:gamma_effect}
	Emissivity of CNM grains for $\gamma=0$ and
	$\gamma=0.1$, where $\gamma$ is the fraction
	of incident H recombining to form H$_2$.
	The H$_2$ formation torques have a minimal effect
	on the emissivity.
	}
\end{figure}

In Figure \ref{fig:j_nu_b} we show the effects of decreasing the
intrinsic dipole moments $\mu_i$
by a factor of two, by taking $\beta_0=0.2\debye$ and $\epsilon=0.05$.
In Figure \ref{fig:gamma_effect} we show the emissivity calculated
for H$_2$ formation efficiencies $\gamma=0$ and 0.1; the
spectra are essentially unchanged.
It is apparent that the emission spectra are not especially
sensitive to the precise values of $\beta_0$ or $\gamma$.

It is evident that the population of small grains which has 
been inferred previously from 
observations of both the UIR emission features and the 12 and 25$\micron$
{\it IRAS} emission 
should produce substantial
electric dipole emission in the 10--100 GHz region.

\begin{figure}
\epsscale{1.00}
\plotone{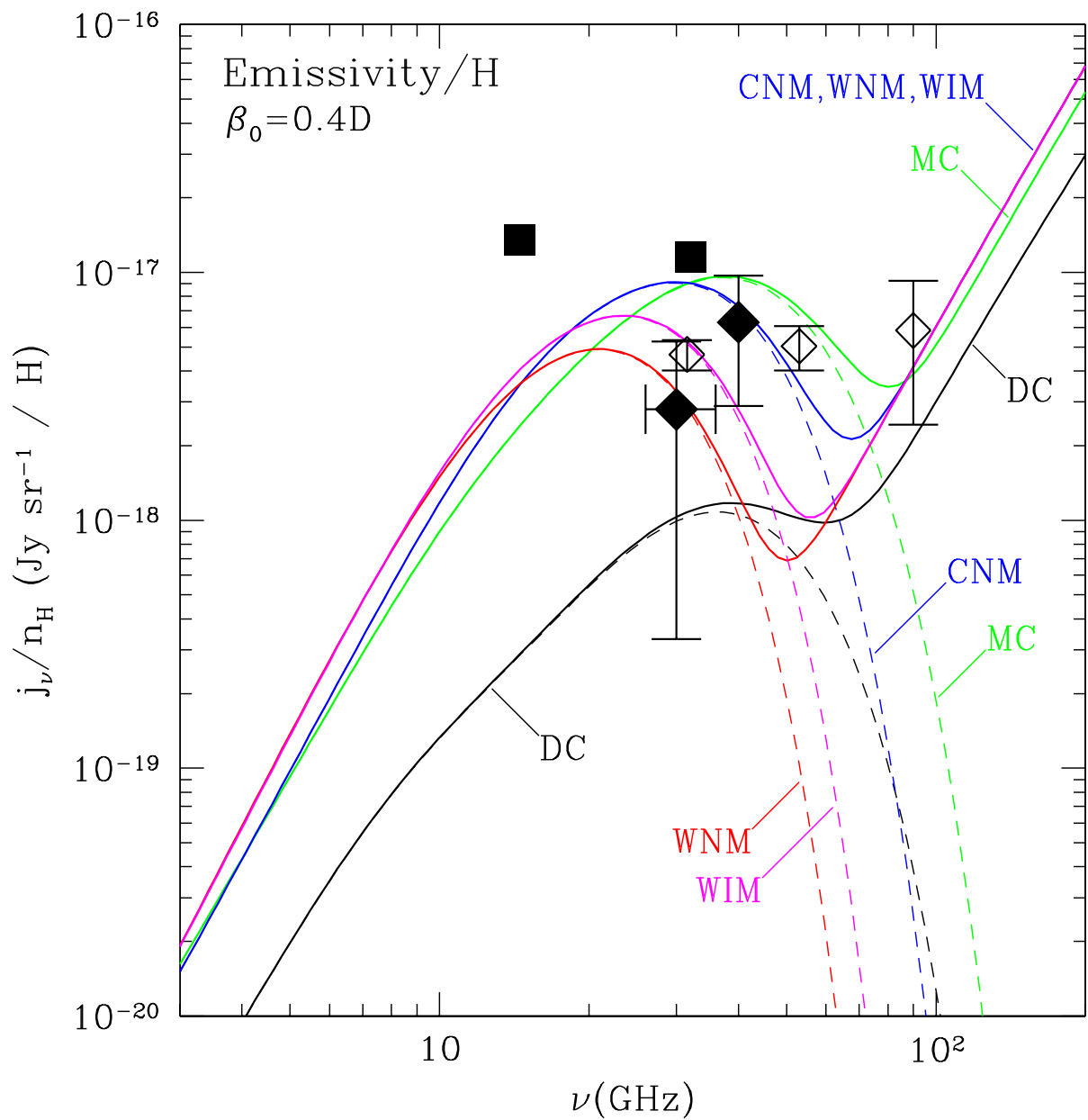}
\caption{
	\label{fig:j_nu}
	Emissivity per H due to ultrasmall 
	spinning dust grains under CNM and WNM
	conditions, for the size distribution of 
	Fig. \protect{\ref{fig:sized}}.
	Also shown are observed emissivities 
	from {\it COBE} DMR (open diamonds; Kogut et al. 1996);
	Saskatoon (filled diamonds; de Oliveira-Costa et al. 1997);
	and OVRO (filled squares; Leitch et al. 1997).
	}
\end{figure}

\begin{figure}
\epsscale{1.00}
\plotone{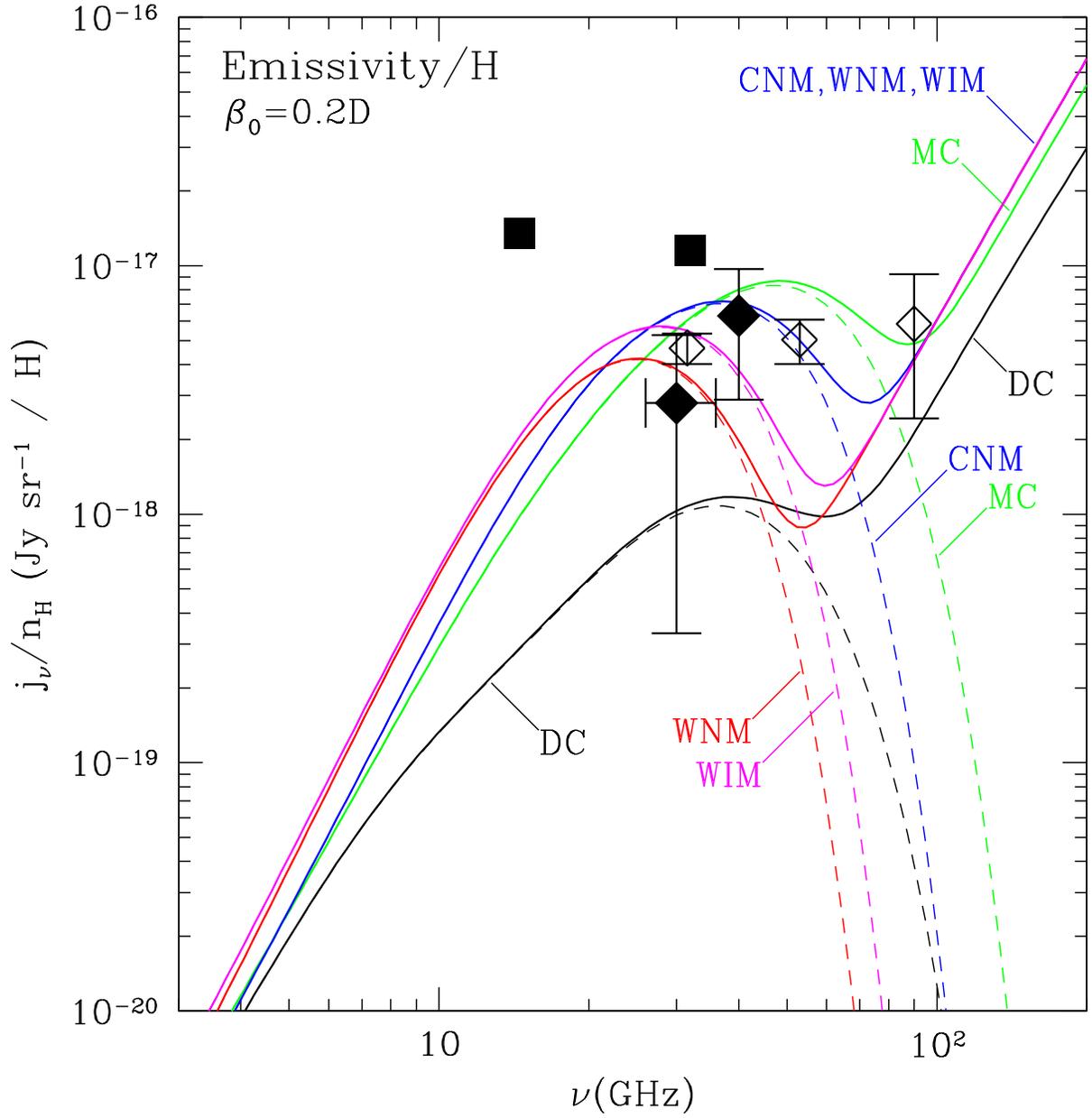}
\caption{
	\label{fig:j_nu_b}
	Same as Fig. \protect{\ref{fig:j_nu}} but for intrinsic
	electric dipole moments $\mu_i$ reduced by a factor of two
	($\beta_0=0.2\debye$).
	}
\end{figure}

It appears that this emission has already been detected by experiments
designed to measure angular structure in the cosmic background
radiation.
Kogut et al. (1996),
de Oliveira-Costa et al (1997),
and
Leitch et al. (1997) have reported detection of $14-90\GHz$ microwave
emission which is correlated with $100\micron$ emission, and therefore
apparently originates in interstellar gas or dust.
Draine \& Lazarian (1998) argue that this emission 
is in fact due to rotational emission from spinning
dust grains in diffuse interstellar gas at high galactic latitudes.
The observed emissivity per H nucleon is shown in 
Figs. \ref{fig:j_nu} -- \ref{fig:j_nu_b},
and it is seen that the observed emissivities are
in fact approximately equal to the electric dipole emission predicted here.
Future measurements of the spectrum of this diffuse 
emission, and its correlation
with interstellar gas and dust, will test this interpretation, although
the emission is weak because of the relatively small amounts
of dust at high galactic latitudes.
The {\it MAP} mission, to be launched in 2000, 
is expected to obtain accurate
maps of the emission from interstellar dust in five bands, from 22 -- 90 GHz.

\section{Detecting Dark Clouds \label{sec:darkcloud}}

It may also be possible to detect emission from spinning dust grains
in dense clouds, where the larger column densities result in increased
intensities.
As an example, we consider the central
surface brightness for the model which Lazarian, Goodman \& Myers (1997)
adopt for the L1755 dark cloud, with a central column density
$N_\H=2\times10^{22}\cm^{-2}$ in material with densities 
$n_\H\approx 10^4 \cm^{-3}$, plus $N_\H=1.3\times10^{21}\cm^{-2}$
in diffuse molecular gas with $n_\H\approx300\cm^{-3}$.
To this we add $N_\H=1\times10^{21}\cm^{-2}$ in HI
surrounding the cloud, with conditions characteristic of the CNM phase.

\begin{figure}
\epsscale{1.00}
\plotone{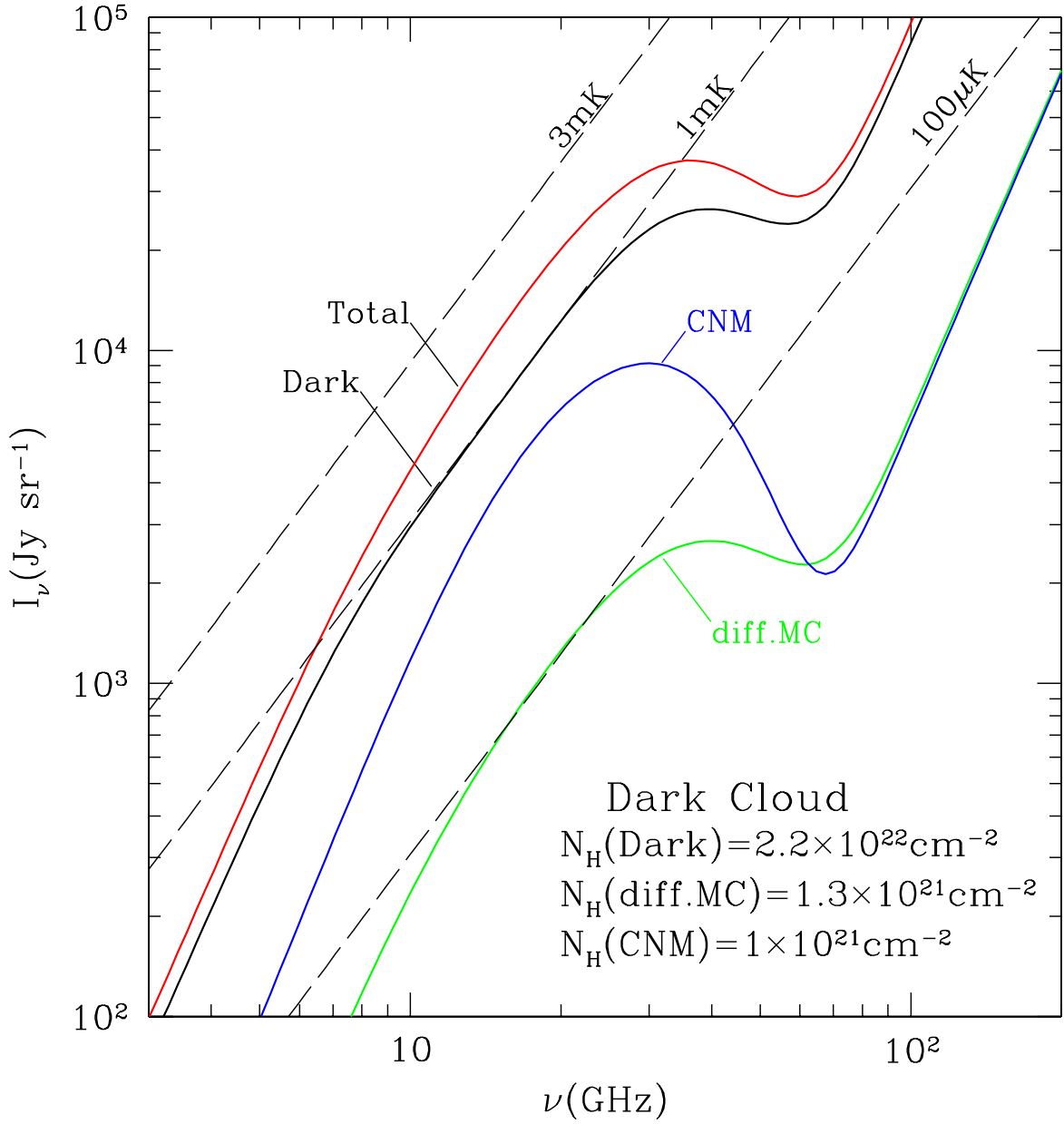}
\caption{
	\label{fig:j_nu_L1755}
	The central surface brightness of dust emission from a
	model cloud resembling L1755 (see text).
	Dashed lines show antenna temperatures of 100$\mu$K and
	0.1mK .
	}
\end{figure}

We add vibrational (``thermal'') emission to the rotational emission,
using dust temperatures of $12\K$, $15\K$, and $18\K$ for dark regions, 
diffuse molecular
gas, and CNM, respectively; the assumed dust opacity is $\propto\nu^{1.7}$
and reproduces the observed $\lambda\gtsim100\micron$ emission from
the interstellar medium for a grain temperature $T_d\approx18\K$.
Figure \ref{fig:j_nu_L1755} shows the resulting spectrum.
We see that at $10 - 30$ GHz 
antenna temperatures of $\sim 1$mK are expected
toward the central regions; for the adopted cloud model
about half of this emission originates in the extended ``CNM'' envelope
of the cloud.

\section{Discussion\label{sec:discussion}}

We have presented above a detailed study of rotational excitation
and damping for grains with sizes less than $\sim10^{-6}\cm$.
In addition to processes included in earlier work by Rouan et al. (1992),
we have included both direct collisions with ions and ``plasma drag''.
Ion collisions and plasma drag,
together with damping by infrared and microwave emission,
dominate rotational excitation and damping for most
interstellar environments.
It is therefore important to include these processes in other studies of
rotational excitation, such as those proposing to explain features
of the diffuse interstellar bands (Rouan, Leger, \& Coupanec 1997).

For the assumed distribution of grain sizes and electric dipole moments,
we predict microwave emission from spinning dust grains which can
account for the ``anomalous emission'' observed recently in the
15--90 GHz range.
The predicted intensities are uncertain, however, as they depend
on three poorly-known factors:
\begin{enumerate}
\item The abundances and size distribution of the small grains.  This
	uncertainty is greatest in the case of dense/dark regions,
	where the absence of starlight denies us the evidence
	(12 -- $25\micron$ emission) which requires an abundance
	of ultrasmall grains in diffuse regions.
\item The charge distribution of the small grains, which affects the
	rates of rotational excitation and damping.
	We have used standard estimates for photoelectric
	cross sections (Bakes \& Tielens 1994) 
	and electron capture cross sections (Draine \& Sutin 1987) for
	very small grains.
	The uncertainties in these quantities
	affect both the electric dipole moment and,
	more importantly, the rate of angular momentum exchange with
	ions.
\item The electric dipole moments of both neutral and charged small grains.
\end{enumerate}
The shape of the small grains (chainlike vs. sheetlike vs. quasispherical)
is also uncertain, but is less critical than the above factors.
A fifth factor --
the efficiency of $\HH$ formation on small grains --
is not critical for these estimates: in Fig.\ \ref{fig:omega} only the
grains in the CNM show appreciable sensitivity
to whether or not $\HH$ formation takes place, and even for the CNM
component the emissivity at $\nu \gtsim 2\GHz$ is only slightly changed
(see Fig.\ \ref{fig:gamma_effect}).

Because of these uncertainties, 
definitive
predictions for the rotational emission spectrum
are not yet possible.
Nevertheless, within the existing theoretical and observational
uncertainties it appears that
much or all of the observed 15--90 GHz ``anomalous'' 
emission is due to spinning dust grains.
The largest discrepancy between observation and theory is at
14.5 GHz, where Leitch et al. report emission about 3.5 times stronger
than the emission predicted in Fig.\ \ref{fig:j_nu}.
Additional measurements at $\nu\ltsim 20\GHz$ will be of great value to
clarify whether this emission has another origin, or whether some of
our assumptions concerning the dust must be modified.

Emission from rotating grains must be allowed for in studies of the
cosmic microwave background radiation (see DL98).
It appears that the small rotating grains may be partially aligned with
the local magnetic field (Lazarian \& Draine 1998), 
so that the electric dipole radiation will
be linearly polarized; this may present a problem for interpretation of
CMB polarization measurements by the MAP mission.

\section{Summary\label{sec:summary}}

The principal results of this paper are as follows:
\begin{enumerate}

\item Even neutral dust grains are expected to usually have electric
dipole moments arising from polarized chemical bonds within the grain.
Charged grains have an additional contribution to the dipole moment
due to displacement of the charge centroid from the mass centroid.
Our estimate for the dipole moment is given by eq.\ (\ref{eq:mu2}).

\item The excitation and damping of rotation in small grains is
determined by collisions with ions and neutrals, 
``plasma drag'', emission of infrared and microwave radiation,
and formation of H$_2$ on the grain surface.
Ion collisions and plasma drag, omitted in previous estimates of
rotation rates, are included in the present analysis and found to often
dominate rotational excitation and damping.
Induced-dipole attraction of neutrals by charged grains, and of ions by
neutral grains, can also be significant.
Because the charge state of the grain, and the fractional ionization of
the gas, do not reflect thermodynamic equilibrium, the fluctuation-dissipation
theorem does not directly apply.
\item For very small grains ($a\ltsim7\times10^{-8}\cm$), the angular
momentum of colliding ions is large compared to the r.m.s. angular momentum
of the grain, and therefore the grain rotation history consists of
``rotational spikes'' separated by intervals of gradual rotational
damping.

\item The estimated grain rotation rates are such that the small grains
which have been postulated to explain the near-infrared emission feature
can account for the emission observed at 30 -- 50 GHz by Kogut et al.,
de Oliveira-Costa et al., and Leitch et al. (1997).

\item The emission observed at 14.5 GHz by Leitch et al. is stronger than
we estimate for spinning grains by a factor $\sim4$.
Additional determinations of
emission from dust at frequencies $\ltsim 20\GHz$ will be of great value.

\item We predict that dark clouds should produce detectable 
microwave emission, with
antenna temperatures of $\gtsim 1 {\rm mK}$ at $\sim 10-30\GHz$.

\end{enumerate}

\acknowledgements
We are grateful to David Spergel for attracting our attention to this problem,
and to Robert Lupton for the availability of the SM package.
We thank W.D. Watson for helpful comments.
B.T.D. acknowledges the support
of NSF grant AST-9619429, and
A.L. the support of NASA grant NAG5-2858.

\newpage
\appendix

\section{Geometric Factors\label{app:geometry}}

\begin{figure}
\epsscale{1.00}
\plotone{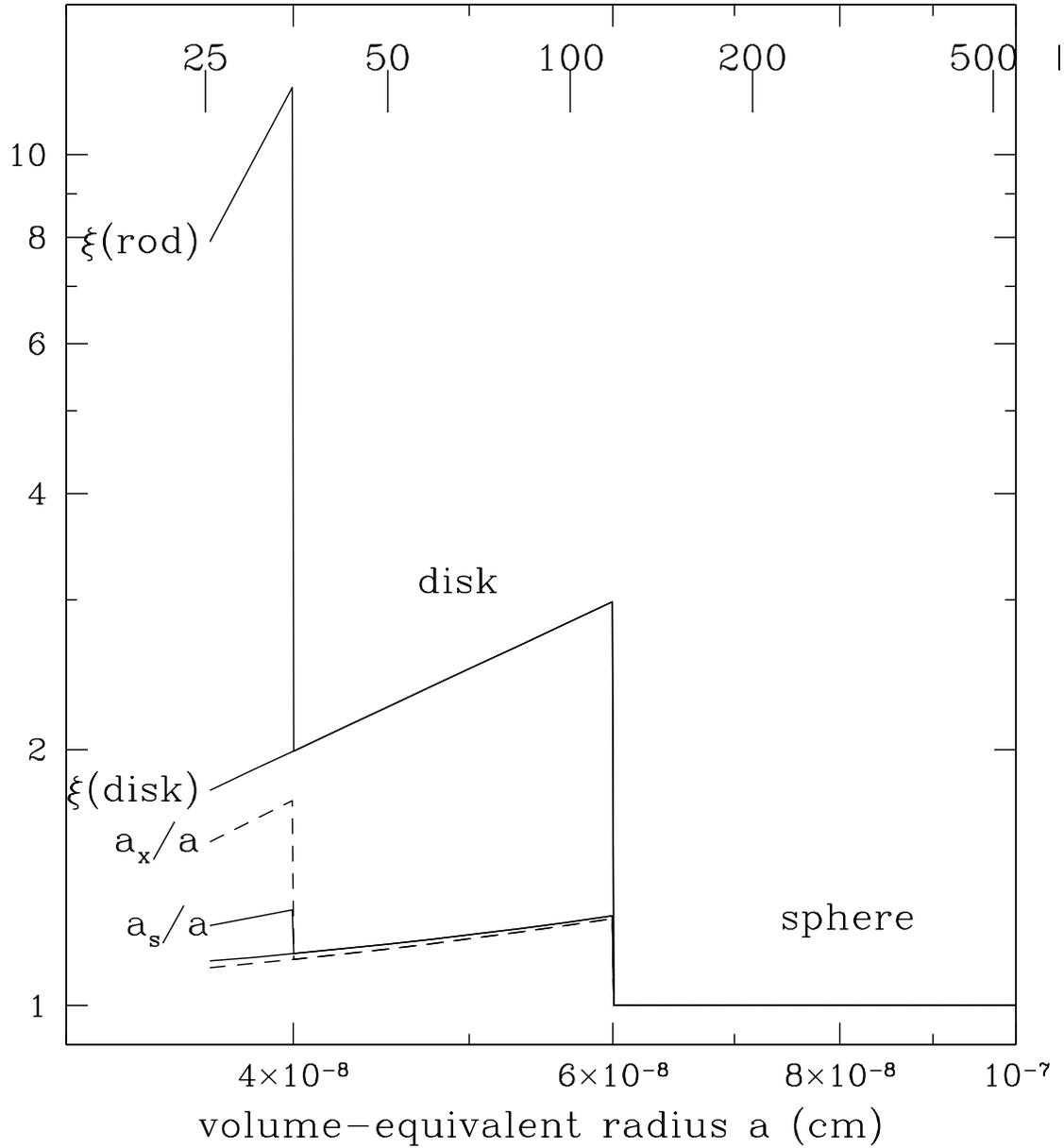}
\caption{
	\label{fig:ax}
	The ratios $a_x/a$ and $\xi$ as a function of $a$.
	Here $a$ is the radius of an equal-volume sphere, $a_x$ is
	the effective size from eq.(\protect{\ref{eq:axdef}},
	and $\alpha$ is the factor by which the moment of inertia
	exceeds that of an equal-volume sphere.
	Tickmarks along the top of the figure show the number of atoms
	per grain.
	}
\end{figure}

For cylindrical grains ($a < a_1$) of diameter $d$ and length $2b$ we have
\beq
b = {8a^3\over 3d^2} ~~~,
\eeq
\beq
a_s = \left[{bd\over2}+ {d^2\over8}\right]^{1/2} ~~~,
\eeq
\beq
a_x = \left[{b^3d\over 6}+{b^2d^2\over 8} + {bd^3\over 8}+{d^4\over64}
\right]^{1/4} ~~~,
\eeq
\beq
\xi = {160\over 27}\left({a\over d}\right)^4 ~~~.
\eeq
where the ``surface-equivalent" radius $a_s$ and ``excitation-equivalent''
radis $a_x$ are defined in eq.(\ref{eq:asdef},\ref{eq:axdef}),
and $\xi \equiv I / (0.4 M a^2)$.
For disklike grains ($a_1 < a < a_2$) of thickness $d$ and radius $b$ we have
\beq
b = \left({4a^3\over 3d}\right)^{1/2} ~~~,
\eeq
\beq
a_s = \left[{b^2\over2} + {bd\over2}\right]^{1/2} ~~~,
\eeq
\beq
a_x = \left[{b^4\over 4} + {b^3d\over 2} + {b^2d^2\over 8} + {bd^3\over24}
\right]^{1/4} ~~~,
\eeq
\beq
\xi = {5a\over 3d} ~~~.
\eeq
For spheres ($a>a_2$), $a_s = a_x = a$ and $\xi=1$.
Figure \ref{fig:ax} shows how $a_s/a$, $a_x/a$ and $\xi$ vary with $a$.

\section{Rotational Damping and Excitation of Grains 
in a Partially Ionized Gas\label{app:rotexc}}

The angular momentum of a spherical grain in a partially-ionized gas
changes due to several distinct scattering processes:
\begin{enumerate}
\item impact of neutral atoms and molecules on the grain surface;
\item impact of ions on the grain surface;
\item electromagnetic interaction of the electric dipole moment of the
grain with passing ions.
\end{enumerate}
Each of these processes contributes to both rotational excitation
and rotational damping.

The microphysics of inelastic scattering of particles which impact directly
on the grain surface is obviously complex.
We idealize the problem by imagining that species which arrive at a point
on the grain surface are re-emitted from the same point on the grain 
surface with a thermal distribution of velocities in the frame of
reference of the local (moving) grain surface at the instant of emission.
Thus we neglect possible effects of large centripetal accelerations of
the grain surface when the grain is rotating rapidly.
We further assume that all arriving neutrals and ions depart the
grain surface as neutrals.

With this idealization, we can now calculate the rate of rotational damping
and rotational excitation.
The drag and rotational excitation from different components is additive.

\subsection{General Considerations\label{app:general}}

Consider a spherical 
grain of radius $a$ and moment of inertia
$I=(8\pi/15)\rho a^5$, with charge $Z_ge$, rotating with angular
velocity $\omega$. 
Atoms and molecules 
arrive at the surface with (on average) zero net angular momentum.
Angular momentum is carried away by the gas atoms and molecules which are
re-emitted from the surface, so the rate of damping is directly
proportional to the rate at which mass arrives at the grain surface.
In order to more easily compare different contributions, we will
``normalize''
to the rate of damping which would be contributed by only the H atoms
in neutral atomic gas:
\beq
{-1\over \omega} {d\over dt} I\omega = 
\left[
n_\H \left({8kT\over \pi m_\H}\right)^{1/2}
\pi a^2 m_\H {2a^2\over 3} 
\right] F ~~~,
\eeq
with $F=1$ for neutral grains and H atoms in atomic gas.
The drag processes are additive, so that
\beq
F = \sum_j F_j
~~~,
\eeq
where $F_j$ is the contribution to $F$ from process $j$.

Suppose that the interaction potential 
is such that particles of species $x$ with
velocity $v$ at infinity collide with the grain surface if and only
if the impact parameter $b \leq b_{max}(v)$.
Then collisions with species $x$ contribute
\beq
F_x = {n_x\over n_\H}\left({m_x\over m_\H}\right)^{1/2}
\left({\pi m_x\over 8 kT}\right)^{1/2}
\int_0^\infty dv ~ 4\pi v^2 f_x(v) v \left({b_{max}(v)\over a}\right)^2~~~,
\eeq
where
\beq
f_x(v)\equiv \left({m_x\over 2\pi kT}\right)^{3/2} \exp(-m_x v^2/kT)
\label{eq:f_x}
~~~.
\eeq

We also consider the rate at which collisions act to increase $J^2$ for
a stationary grain, and again normalize to the rate which would be
appropriate for H atoms in neutral atomic gas 
if the atoms left the grain as H atoms at temperature $T$:
\beq
\left\langle {d\over dt} J^2\right\rangle_{\omega=0}
=
\left[ 
n_\H \left({8kT\over \pi m_\H}\right)^{1/2} 
4\pi a^4 m_\H kT 
\right]
G
~~~.
\eeq
For a given grain, we will consider the contributions $G_j$ which
various processes $j$ make to the overall excitation rate $G$.
The incoming particles contribute
\beq
G_x^{(in)} =
{n_x\over n_\H}
\left({m_x\over m_\H}\right)^{1/2}
{\pi^{1/2}\over 8}
\left( {m_x\over 2kT}
\right)^{3/2}
\int_0^\infty dv ~4\pi v^2 f_x(v) v^3
\left( {b_{max}(v)\over a}
\right)^4
~~~.
\eeq
Each arriving particle, after delivering its incoming angular momentum
to the grain, later ``evaporates'' and escapes from the grain surface.
We assume that evaporating particles have a velocity distribution
appropriate for a grain temperature $T_{ev}$.
Because the evaporating particles interact with the grain (some
evaporating particles will not have ``escape velocity'' and will
return to the grain surface),
we use the fact that the angular momentum distribution
of escaping particles must equal that for the same species being
captured from a gas at temperature $T_{ev}$.
Thus one can show that the evaporating particles contribute
\beq
G_x^{(ev)} =
{m_x \over 8 a^2 kT} \Delta F
\left\{
{
\int_0^\infty dv f_x^{(ev)}(v) v^5 \left[ b_{max}^{(ev)}(v)\right]^4
\over
\int_0^\infty dv f_x^{(ev)}(v) v^3 \left[ b_{max}^{(ev)}(v)\right]^2
}
\right\}~~~,
\eeq
where $f_x^{(ev)}(v)$ is given by eq.(\ref{eq:f_x}) but with
$T$ replaced by $T_{ev}$,
and $b_{max}^{(ev)}(v)$ is the value of $b_{max}$ appropriate for
particles with the physical properties of the evaporating species
(i.e., neutrals even when the impinging species $x$ is an ion).

\subsection{Neutral-Grain Collisions\label{app:F_n}}

Prior to impact, the grain-neutral interaction is approximated by
an induced dipole potential,
\beq
U(r) = - {1\over 2}\alpha_n {Z_g^2e^2\over r^4}
\eeq
where $\alpha_n$ is the polarizability of the
neutral atom or molecule, and $Z_ge$ is the charge on the grain.
The trajectories of particles in an $r^{-4}$ potential are well-known
(see, e.g., Wannier 1953, Osterbrock 1961).
For an atom with initial velocity $v$, 
impact parameters 
\beq
b < b_0(v)\equiv \left({4Z_g^2 e^2\alpha_n \over m v^2}\right)^{1/4}
\eeq
have ``spiral'' trajectories which pass through the origin, 
while for $b > b_0$ trajectories are
``hyperbolic'', with a finite distance of closest approach.
A hyperbolic trajectory with $b=b_0$ has a distance of closest
approach $r_{min}=b_0/\surd2$.
Thus we may use the condition $a = b_0/\surd2$ to define a velocity 
\beq
v_a \equiv \left( {Z_g^2 e^2 \alpha_n \over m a^4}\right)^{1/2}
\eeq
such that 
\beq
b_{max}(v) =
\left\{
\begin{array}{ll}
	b_0(v) & \mbox{if $v \leq v_a$} \\
	a\left(1+ {Z_g^2e^2\alpha_n\over m a^4v^2}\right)^{1/2}
	&\mbox{if $v \geq v_a$}
\end{array}
	\right.
\eeq
Thus
the contribution to $F$ is
\beq
F_n = 
{n_n\over n_\H}\left({m_n\over m_\H}\right)^{1/2}
\left[\exp(-\epsilon_n^2) + \epsilon_n\pi^{1/2}{\rm erf}(\epsilon_n)\right]
\label{eq:deltaf1n}
\eeq
where
\beq
\epsilon_n^2 \equiv {mv_a^2\over 2kT} = {Z_g^2e^2\alpha_n\over 2a^4kT}
~~~.
\eeq
The contribution of neutral species $n$ to $G$ is
\beq
G_n = \left({T+T_{ev}\over 2T}\right)
{n_n\over n_\H}\left({m_n\over m_\H}\right)^{1/2}
\left[\exp(-\epsilon_n^2) + 2 \epsilon_n^2\right]
\label{eq:deltaf2n}~~~.
\eeq

\subsection{Ion-Grain Collisions\label{app:F_i}}

\subsubsection{Charged Grain: $Z_g\neq0$\label{app:F_i_Z}}

For a Coulomb law potential $Z_gZ_ie^2/r$, we have (Spitzer 1941)
\beq
b_{max}(v) = 
\left\{
\begin{array}{ll}
0 & \mbox{for ${mv^2\over 2} < {Z_g Z_i e^2\over a}$}\\
a\left(1-{2Z_g Z_i e^2\over mav^2 }\right)^{1/2} &
\mbox{for ${mv^2\over 2} > {Z_g Z_i e^2\over a}$}
\end{array}
\right.
\eeq
For ions and charged grains ($Z_gZ_i\neq0$)
we neglect the modifications to $b_{max}$ which result when
the ``image charge'' contribution to the interaction is taken into
consideration (see Draine \& Sutin 1987, and \S\ref{app:F_i_0} below).
The contribution of colliding ions to $F$ is then
(Spitzer 1941)
\beq
F_i = 
{n_i\over n_\H} 
\left(	
	{m_i\over m_\H}
\right)^{1/2} 
g_1(\psi)
~~~~\mbox{for $Z_iZ_g\neq 0$}~~~,
\eeq
\beq
\psi\equiv{Z_g Z_i e^2\over akT}~~~,
\eeq
\beq
g_1(x) = \left\{
\begin{array}{ll}
	1-x	& \mbox{if $x < 0$} \\
	e^{-x} & \mbox{if $x > 0$}
\end{array}
	\right.~~~.
\eeq
The contribution to $G$ due to the arriving ions is
(Anderson \& Watson 1993)
\beq
G_i^{(in)} =
{1\over2}{n_i\over n_\H}
\left( {m_i\over m_\H}\right)^{1/2}
g_2(\psi)~~~,
\eeq
\beq
g_2(x) = \left\{
\begin{array}{ll}
	1-x+x^2/2	& \mbox{if $x < 0$} \\
	e^{-x} & \mbox{if $x > 0$}
\end{array}
	\right.~~~.
\eeq
We assume the ions to depart from the grain as neutrals.  Per
departing neutral, we should have the same contribution to $G$
as for arriving neutrals if the gas temperature were $T_{ev}$ (see
eq. \ref{eq:deltaf1n} and \ref{eq:deltaf2n}).
Thus
departing neutrals contribute
\beq
G_i^{(ev)} =  F_i
{T_{ev}\over 2T}
\left[
{
\exp(-Z_g^2\epsilon_i^2) + 2Z_g^2\epsilon_i^2
\over
\exp(-Z_g^2\epsilon_i^2) +|Z_g|\epsilon_i \pi^{1/2} {\rm erf}(|Z_g|\epsilon_i)
}
\right]~~~,
\eeq
where
\beq
\epsilon_i^2 \equiv {e^2 \alpha_i \over 2 a^4 kT_{ev}}~~~,
\eeq
where $\alpha_i$ is the polarizability of the neutral obtained from
ion $i$.
\subsubsection{Neutral Grain: $Z_g=0$\label{app:F_i_0}}

In the case of a neutral grain, the collision rate is modified by
the polarization of the grain by the electric field of the ion.
If the grain is approximated as a perfect conductor, the interaction
potential is
\beq
U(r) = -{Z_i^2 e^2 a^3 \over 2r^2(r^2-a^2)}~~~.
\eeq
The collision rate for this interaction potential has 
been discussed by Draine \& Sutin (1987).
Using their eq. (B6),\footnote{
	Note the typographical error in eq. (B6) of
	Draine \& Sutin (1987), which should read
	$(2\epsilon x-\nu)(x^2-1)^2 -x=0$.
	}
with $\nu=0$, gives a critical impact parameter
\beq
b_{max}(v) 
= a\left[1 + \left({4Z_i^2e^2\over m v_0^2 a}\right)^{1/2}\right]^{1/2}~~~,
\eeq
so that
\beq
F_i = {n_i\over n_\H}\left({m_i\over m_\H}\right)^{1/2}
\left[ 1 +  {\pi^{1/2}\over 2}\phi \right]~~~.
\eeq
The arriving ions contribute
\beq
G_i^{(in)} = {1\over2}{n_i \over n_\H} \left({m_i \over m_\H}\right)^{1/2}
\left[ 1 + {3\pi^{1/2}\over4}\phi + {1\over2}\phi^2\right]~~~,
\eeq
where
\beq
\phi^2\equiv{2 Z_i^2 e^2 \over a k T}~~~,
\eeq
and the departing neutrals contribute an additional
\beq
G_i^{(ev)} = {T_{ev}\over 2T}\Delta F~~~.
\eeq

\subsection{Plasma Drag\label{app:F_p}}

Consider a stationary grain with electric dipole moment $\bmu$.
The electric field $\bE$ at the grain due to nearby ions exerts
a torque $\bmu\times\bE$.
To estimate the effects of such torques, we assume that an
ion with impact parameter $b$ passes on a straight-line trajectory at 
constant velocity $\bv$.
To define the orientation of $\bmu$, let the direction of 
$\bv$ define a polar axis,
with $\theta$ the angle between $\bv$ and $\bmu$, and let the
azimuthal angle $\phi=0$ when $\bmu$ is in the plane containing
both the ion trajectory and the grain.
It is then easy to show that the angular momentum $\Delta J$
exchanged between the ion and the grain has
\beq
(\Delta J)^2 = \left({2\mu Z_i e \over b v}\right)^2
\left( \sin^2\theta\sin^2\phi + \cos^2\theta \right)~~~.
\eeq
If we now average over random orientation of $\bmu$
($\langle \sin^2\theta\rangle=2/3$, $\langle \cos^2\theta\rangle=1/3$,
$\langle\sin^2\phi\rangle=1/2$) and then over a thermal distribution
of velocities $v$, we find (Anderson \& Watson 1993)
\begin{eqnarray}
{d\over dt} J^2 &=& n_i \int_0^\infty dv ~4\pi v^2 f_i(v) v \int_{b_1}^{b_2}
2\pi b db ~{2\over 3}\left({2\mu Z_i e \over b v}\right)^2\\
\nonumber
&=& 
{8\pi\over3} n_i \left({8kT\over\pi m_i}\right)^{1/2} 
{Z_i^2 e^2 m_i \over kT} \mu^2 \ln(b_2/b_1)~~~.
\label{eq:gpderiv}
\end{eqnarray}
For the lower cutoff we simply take the 
grain radius $b_1=a$.
The upper cutoff is more problematic.
It is clear that it cannot exceed
the Debye length,
\beq
b_2 \leq \lambda_D \equiv \left( {kT\over 4\pi n_e e^2}\right)^{1/2}
= 398 \left({T /100\K \over n_e/0.03\cm^{-3}}\right)^{1/2} \cm~~~,
\eeq
since at larger distances the ion electric field is screened.
However, for a grain
rotating at angular velocity $\omega >0$, an ion with
velocity $v_{th}=(2kT/m_i)^{1/2}$
and impact parameter 
$b \gtsim b_\omega\equiv v_{th}$
will be almost unaffected by the rotating 
component of the grain dipole moment,
as the torque averaged over the grain rotation will
tend to zero.
For a grain with a thermal rotation rate $\omega \approx (2kT/I)^{1/2}$
we have
\beq
b_\omega \approx \left({8\pi\xi\rho a^5\over 15 m_i}\right)^{1/2}
= 4.5 \times10^{-6}a_{-7}^{2.5} \xi^{1/2}\left({m_\H\over m_i}\right)^{1/2}
\cm ~~~.
\eeq
Let $\Psi$ be the angle between the dipole moment $\bmu$ and the rotation
velocity $\bomega$.
In eq.(\ref{eq:gpderiv}) we would then replace
\beq
\mu^2\ln(b_2/b_1) \rightarrow 
\mu^2\left[\ln(b_\omega/b_1)+
\cos^2\Psi\ln(b_2/b_\omega)\right]
\eeq
We differ here from the treatment of Anderson \& Watson (1993)
who neglected the contribution of impact parameters $b>b_\omega$.
The rotational excitation is quantized,
and we expect the above classical estimate to fail when the characteristic
frequency $\sim v_{th}/b$ of the time-varying electric field
from the passing ion varies falls below the frequency $\hbar/I$ of the
$J=0\rightarrow1$ rotational transition, resulting in a ``quantum''
cutoff $b_q\approx I v_{th}/\hbar$, or
\beq
b_q = 4.1\times10^{-3}\xi a_{-7}^5 
\left({T/100\K\over m_i/m_\H}\right)^{1/2}\cm ~~~.
\eeq
Passing ions then contribute\footnote{
	It is instructive to compare \ref{eq:dj2dt_p}
	with quantum-mechanical calculations.
	For CN ($\mu=1.45\debye$) and $T=10^4\K$ the rate coefficient for
	$J/\hbar=0\rightarrow1$ excitation by protons is 
	$\langle\sigma v\rangle_{0\rightarrow1}
	\approx4\times10^{-6}\cm^3\s^{-1}$
	(Thaddeus 1972), whereas taking $\cos^2\Psi=1$ we estimate 
	$(2n_i \hbar^2)^{-1}dJ^2/dt \approx 1.5\times10^{-5}\cm^3\s^{-1}$,
	too large by a factor $\sim4$.
	Such agreement even for a case where quantum effects are dominant
	is reassuring.
	}
\beq
{d J^2 \over dt} = {16 \over 3} n_i 
\left( {2 \pi m_i \over kT}\right)^{1/2}  
{Z_i^2 e^2}
\mu^2
\left\{ \ln (b_\omega/a_s) + \cos^2\Psi 
\ln \left[ \min(b_q,\lambda_D)/b_\omega \right] \right\}~~~,
\label{eq:dj2dt_p}
\eeq
\beq
G_p = {n_i \over n_\H}\left( {m_i \over m_\H} \right)^{1/2}
{2Z_i^2 e^2 \over 3 a_x^4 (kT)^2}\mu^2
\left\{ \ln (b_\omega/a_s) + \cos^2\Psi 
\ln \left[ \min(b_q,\lambda_D)/b_\omega \right] \right\}~~~.
\label{eq:app:G_p}
\eeq
Having estimated the excitation rate $G_p$,
we may now invoke
the fluctuation-dissipation theorem to see that the normalized
contribution to the damping rate must be simply
$F_p = G_p$, so that if torques from passing ions were the
only torques present, the grain would attain a thermal 
rotational distribution.

\section{Damping and Excitation of Rotation by IR Emission\label{app:IRdamping}}

\subsection{Steady Emission\label{app:IR,c}}

Suppose that the grain at rest contains six rotating electric dipoles,
each with electric dipole moment $p$ and
with angular velocity $\omega$ in grain body coordinates.
There is one dipole rotating clockwise and one counterclockwise
around each of the $x-$, $y-$, and 
$z-$axes.
The dipoles emit incoherently, so that at rest the
power radiated is just
\beq
{dE\over dt} = 6\times{2\omega^4 p^2\over 3c^3} ~~~;
\eeq
the emission is isotropic and unpolarized.
Each dipole emits radiation which, in directions parallel to the dipole's
rotation axis, is 100\% circularly polarized.
For an individual dipole the angular momentum loss $dL^{(1)}/dt$ 
is just
\beq
\frac{dJ^{(1)}}{dt}={2\omega^3 p^2 \over 3 c^3} ~~~.
\eeq

A system of two oppositely rotating dipoles radiates a net angular
momentum if the whole system 
rotates with frequency $\omega_{\rm r}$ along the dipoles' rotation
axis:
in the rest frame 
the frequencies of emission for the two dipoles
are, respectively, $\omega+\omega_{\rm r}$ and $\omega-\omega_{\rm r}$.

Let the system of six dipoles rotate around the $x$-axis 
with frequency $\omega_{\rm r}\ll \omega$. The rate of loss
of angular momentum associated with the single emission frequency
$\omega \gg \omega_r$ is then
\beq
\frac{dJ}{dt}=\left[\frac{2}{3}\frac{(\omega+\omega_{\rm r})^3}{c^3}-
\frac{2}{3}\frac{(\omega-\omega_{\rm r})^3}{c^3}\right]p^2
\approx\frac{\omega_{\rm
r}}{\omega^2}\frac{dE}{dt}~~~.
\eeq

Using the Planck expression for thermal emissivity we find
\beq
\frac{dJ}{dt}=
-\omega_{\rm r}a^2\int^{\infty}_0\frac{Q_{\nu}B_{\nu}}{\nu^2}
d\nu ~~~,
\eeq
while
\beq
{dE\over dt} =
-4\pi^2 a^2\int_0^\infty Q_\nu B_\nu d\nu ~~~.
\label{eq:dEdt}
\eeq
Now suppose 
\beq
Q_{\nu}=Q_0 \left(\frac{\nu}{\nu_0}\right)^{\beta}~~~,
\eeq
where we will assume $\beta=2$ for interstellar dust (Draine \& Lee 1984),
and set
\beq
{dE\over dt} =
-\pi a^2 \langle Q\rangle_* u_* c ~~~,
\eeq
where $u_*$ is the energy density of starlight, and $\langle Q\rangle_*$
is the grain absorption efficiency averaged over the starlight spectrum.
Then
\beq
{dJ\over dt} = -{h^2\over 4\pi}{\Gamma(\beta+2)\over\Gamma(\beta+4)}
{\zeta(\beta+2)\over\zeta(\beta+4)}
{\langle Q\rangle_* a^2 u_* c \over (kT_d)^2}
\label{eq:dJdt}~~~,
\eeq
where $\Gamma$ and $\zeta$ are the usual gamma function and 
Riemann $\zeta$-function.
Thus
\begin{eqnarray}
F_{\IR,c} 
&=& {3\over (8\pi)^{3/2}}
{\Gamma(\beta+2)\zeta(\beta+2)\over \Gamma(\beta+4)\zeta(\beta+4)}
{\langle Q\rangle_* u_* c h^2 \over n_\H (m_\H kT)^{1/2} (a k T_d)^2}
\\
&=&
{59.0\over a_{-7}}
\left({u_*\over u_\ISRF}\right)
\left({20\cm^{-3}\over n_\H}\right)
\left({100\K\over T}\right)^{1/2}
\left({20\K\over T_d}\right)^2 ~~~.
\end{eqnarray}
The grain temperature $T_d$ and starlight intensity $u_*$ are
related through
\beq
T_d = {hc\over k}\left[ {\langle Q\rangle_* u_* \over
8\pi h c Q_0 \lambda_0^\beta\Gamma(\beta+4)\zeta(\beta+4)}
\right]^{1/(\beta+4)}
\eeq
where $\lambda_0=c/\nu_0$.
We consider three grain materials: graphite, ``astronomical silicate",
and $\alpha-$SiC (Draine \& Lee 1984; Laor \& Draine 1993), for each
of which the infrared emissivity has $\beta=2$, so that $T_d\propto u_*^{1/6}$.

The recoil from photon emission is a source of rotational excitation.
For a nonrotating grain with radius $a\ll hc/kT_d$, 
\beq
\frac{d}{dt}J^2 = \frac{dN_{\rm ph}}{dt}\hbar^2
\label{dJ}~~~,
\eeq
where the photon emission rate is
\beq
{dN_{\rm ph}\over dt} =
{\Gamma(\beta+3)\over\Gamma(\beta+4)}{\zeta(\beta+3)\over\zeta(\beta+4)}
{1\over kT_d}\pi a^2 \langle Q\rangle_* u_* c ~~~.
\eeq
Thus
\beq
G_{\IR,c}
=
{\langle Q\rangle_* u_* h^2 c \over
16 n_\H m_\H^{1/2} (2\pi kT)^{3/2} a^2 kT_d}
{\Gamma(\beta+3)\zeta(\beta+3)\over\Gamma(\beta+4)\zeta(\beta+4)} ~~~.
\eeq

\subsection{Thermal Spikes\label{app:IR,q}}

Grains undergo a sudden temperature rise following each photon
absorption.
For very small grains, the temperature history can be regarded as a
sequence of independent ``thermal spikes'', separated by intervals
where the grain is very cold.
Rouan et al. (1992) discussed the rotational damping for 
a vibrational density of states chosen to approximate a PAH molecule.
Here we adopt the Debye model for the 
%grain's heat capacity 
heat capacity of the $3N-6$ vibrational modes 
(Kittel 1972).
If $\Theta$ is the Debye temperature,
%and $n$ is the atomic number density in the solid, 
the thermal energy content for $T\ll\Theta$ is
\beq
E=
%{4\pi^5\over 5}a^3 n k T \left(\frac{T}{\Theta}\right)^3~~~.
{3\pi^4\over 5}(N-2) k T \left(\frac{T}{\Theta}\right)^3~~~.
\label{eq:EofT}
\eeq
The absorbed energy of a UV photon first heats the grain
and then is emitted in the infrared.
From (\ref{eq:dEdt}) and (\ref{eq:dJdt}) we obtain
\beq
{dJ\over dE} = {\omega_r\over 4\pi^2}{h^2\over (kT)^2}
{\Gamma(\beta+2)\zeta(\beta+2)\over\Gamma(\beta+4)\zeta(\beta+4)}~~~.
\eeq
Using (\ref{eq:EofT}) to eliminate $T$ in favor of $E$, we
integrate to obtain the angular momentum loss in cooling from
initial energy $E$ to $T=0$:
\beq
\delta J= -A\omega_r E^{1/2}~~~,
\eeq
\beq
A\equiv 
%h^2\left({\pi\over5}\right)^{1/2}
%\frac{\Gamma(\beta+2)\zeta(\beta+2)}{\Gamma(\beta+4)\zeta(\beta+4)}
%{n^{1/2}a^{3/2}\over (k\Theta)^{3/2}}~~~.
%
{h^2\over 2}\left({3\over5}\right)^{1/2}
\frac{\Gamma(\beta+2)\zeta(\beta+2)}{\Gamma(\beta+4)\zeta(\beta+4)}
{(N-2)^{1/2}\over (k\Theta)^{3/2}}~~~.
\eeq
The energy $E$ is provided by the interstellar radiation field.
Therefore
\beq
\frac{dJ}{dt}=-A \omega_{\rm r}\int^{\infty}_0 d\nu
\left(\frac{c u_{\nu}}{h\nu}\right)Q_{\nu}
\left(h\nu\right)^{1/2}\pi a^2~~~.
\eeq

Define
\beq
\calF\equiv \int^{\infty}_0 d\nu \frac{c u_{\nu}}{h\nu}~~~,
\label{eq:calF}
\eeq
\beq
\langle E\rangle_\calF
\equiv
{1\over\calF}\int d\nu ~cu_\nu~~~,
\eeq
\beq
\langle Q \rangle_\calF
\equiv
\frac{1}{\calF}\int^{\infty}_0 d\nu
\left(\frac{c u_{\nu}}{h\nu}\right)Q_{\nu}~~~,
\label{eq:Q_calF}
\eeq
\beq
\langle Q \rangle_{\calF E}
\equiv
\frac{1}{\calF}\int^{\infty}_0 d\nu
~ c u_{\nu} ~Q_{\nu}~~~,
\label{eq:Q_calFE}
\eeq
\beq
\langle E^\gamma \rangle_{\calF Q}
\equiv
\frac{1}{\calF \langle Q \rangle_\calF}
\int^{\infty}_0 d\nu
\left(\frac{c u_{\nu}}{h\nu}\right)Q_{\nu}
\left(h\nu\right)^\gamma~~~.
\label{eq:Egamma_calFQ}
\eeq
Then
\beq
\frac{dJ}{dt}=-A \omega_{\rm r}
\calF
\langle Q \rangle_\calF 
\langle E^{1/2} \rangle_{\calF Q} \pi a^2~~~.
\eeq

Thus
\beq
F_{\IR,q}=
%{3\pi\over 4(10)^{1/2}} 
%{\Gamma(\beta+2)\zeta(\beta+2)\over\Gamma(\beta+4)\zeta(\beta+4)}
%{h^2 n^{1/2}\over a^{1/2}(k\Theta)^{3/2}}
%
{3\over 8}\left({3\pi\over 10}\right)^{1/2}
{\Gamma(\beta+2)\zeta(\beta+2)\over\Gamma(\beta+4)\zeta(\beta+4)}
{h^2 (N-2)^{1/2}\over a^2(k\Theta)^{3/2}}
{
\calF \langle Q\rangle_\calF \langle E^{1/2}\rangle_{\calF Q} \over
n_\H (m_\H kT)^{1/2}
}~~~.
\label{eq:firq}
\eeq
Because $\langle Q\rangle_\calF\propto a$, 
$F_{\IR,q}\propto (N-2)^{1/2}/a$.

The emission following each heating event also contributes to the
rotational excitation of the grain, with
\beq
\Delta J^2 = 
{h^2\over 3\pi}
%\left({4\pi\over3}\right)^{1/4}
{\Gamma(\beta+3)\zeta(\beta+3)\over\Gamma(\beta+4)\zeta(\beta+4)}
%n^{1/4} a^{3/4}
%
(N-2)^{1/4}
\left( {E\over k\Theta}\right)^{3/4} ~~~.
\label{eq:dJ^2}
\eeq
Thus
\beq
G_{IR,q} =
\left[
{h^2\over 3\pi}
%\left({4\pi\over3}\right)^{1/4}
{\Gamma(\beta+3)\zeta(\beta+3)\over\Gamma(\beta+4)\zeta(\beta+4)}
%{n^{1/4} a^{3/4}\over (k\Theta)^{3/4}}
{(N-2)^{1/4}\over (k\Theta)^{3/4}}
\right]
{\calF\langle Q\rangle_\calF \langle E^{3/4}\rangle_{\calF Q}
\over
n_\H (8kTm_\H/\pi)^{1/2} 4a^2kT}~~~.
\eeq
$\langle Q \rangle_\calF$,
$\langle E^{1/2} \rangle_{\calF Q}$,
and
$\langle E^{3/4} \rangle_{\calF Q}$
were calculated for small graphite and silicate spheres for
the interstellar radiation spectrum 
of Mezger, Mathis, \& Panagia (1982)
and Mathis, Mezger, \& Panagia (1983).
The results are presented in Table \ref{tab:FQEcalcs}

\clearpage
%-------------------- begin table ------------------------
\begin{table}
\begin{center}
\caption{
	Idealized phases for interstellar matter.
	\label{tab:phases}
	}
\begin{tabular}{ | c | c| c | c | c | c | }
\hline
phase			&DC	&MC	&CNM	&WNM	&WIM	\\
\hline
$n_\H (\cm^{-3})$	&$10^4$	&300.	&30	&0.4	&0.1	\\
$T (\K)$		&10.	&20.	&100.	&6000.	&8000.	\\
$\chi$			&$10^{-4}$&0.01	&1.	&1.	&1.	\\
$x_\H\equiv n(\H^+)/n_\H$ &0	&0	&0.0012	&0.1	&0.99	\\
$x_M\equiv n(M^+)/n_\H$	&$10^{-6}$&0.0001&0.0003	&0.0003	&0.001	\\
$y\equiv 2n(\HH)/n_\H$	&0.999	&0.99	&0.	&0.	&0.	\\
\hline
\end{tabular}
\end{center}
\end{table}
%-----------------------end table---------------------------

%------------------------ begin table ------------------
\begin{table}
\begin{center}
\caption{
	Electric Dipole Moments for Selected Bonds$^a$
	\label{tab:bonds}
	}
\begin{tabular}{ | c| c | }
\hline
bond					&$\mu$	\\
\hline
aliphatic C--H		&0.3	\\
aliphatic C=O		&2.4	\\
aromatic C=O		&2.65	\\
aromatic C-OH		&1.6	\\
aliphatic C-OH		&1.7	\\
aromatic C-CH$_3$	&0.37	\\
aromatic C-C$\equiv$CH	&0.7	\\
aliphatic C-C$\equiv$CH	&0.9	\\
aromatic C-CHO		&2.96	\\
aliphatic C-CHO		&2.49	\\
\hline
\end{tabular}
\end{center}
$^{a}$ Dean (1992)
\end{table}
%------------------------------- end table ---------------------
%----------------------begin table------------------------------
\begin{table}
\begin{center}
\caption{
	Electric Dipole Moments for Selected Molecules
	\label{tab:betas}
	}
\begin{tabular}{ | c| c | c | c | }
\hline
molecule				& $N$	&$\mu$	&$\beta$\\
\hline
C$_5$H$_8$ 1 pentyne (trans)$^{a}$	&13	&0.84	&0.23	\\
C$_7$H$_5$N benzonitrile$^{a}$		&13	&4.18	&1.16	\\
HC$_{11}$N cyanopolyyne$^{b}$		&13	&5.47	&1.52	\\
C$_5$H$_{10}$O cyclopentanol$^{a}$	&16	&1.72	&0.43	\\
C$_8$H$_{10}$ ethylbenzene$^{a}$	&18	&0.59	&0.14	\\
C$_8$H$_{10}$ orthoxylene$^{a}$		&18	&0.62	&0.15	\\
C$_{12}$H$_{10}$ acenaphthene$^{a}$	&22	&0.85	&0.18	\\
\hline
\end{tabular}
\end{center}

$^{a}$ CRC Handbook (1997)

$^{b}$ Bell et al. (1997)

\end{table}
%------------------------ end table --------------------
%-------------------------begin table---------------------------
\begin{table}
\begin{center}
\caption{
	Dust Grain Properties for Interstellar Starlight
	\label{tab:FQEcalcs}
	}
\begin{tabular}{|c|c|c|c|}
\hline
quantity   & graphite     & silicate & $\alpha$-SiC \\ 
\hline 
$\langle Q \rangle_\calF/a_{-7}$	&$3.12\times10^{-3}$	&$5.79\times10^{-4}$
							&$9.06\times10^{-4}$\\
$\langle Q \rangle_{\calF E}/a_{-7}$	&$8.92\times10^{-3}$	&$3.20\times10^{-3}$
							&$8.45\times10^{-3}$\\
$\langle E^{1/2} \rangle_{\calF Q}^2$ 	& 2.56~eV	& 4.63~eV
							& 9.41~eV\\
$\langle E^{3/4}\rangle_{\calF Q}^{4/3}$& 2.78~eV	& 5.28~eV
							& 9.68~eV\\
$\langle E \rangle_{\calF Q}$		& 3.03~eV	& 5.86~eV
							& 9.87~eV\\
$Q(100\micron)/a_{-7}$			&$1.93\times10^{-3}$	&$1.44\times10^{-5}$
							&$9.03\times10^{-7}$\\
$T_d(\ISRF)$			& 19.7~K		& 17.4~K
							& 32.5~K\\
\hline
\end{tabular}
\end{center}
Assuming interstellar starlight (Mathis, Mezger, \& Panagia 1982;
	Mezger, Mathis \& Panagia 1983)
	with
	$\calF=1.525\times 10^{10}\cm^{-2}\s^{-1}$, 
	$u_\ISRF= 8.626 \times 10^{-13}\erg\cm^{-3}$, and
	$\langle E \rangle_\calF= 1.058\eV$.
	See eq.(\protect{\ref{eq:calF}}--\protect{\ref{eq:Q_calFE}}).
	
\end{table}
\clearpage
%--------------------------------------------------------

\pagebreak

%--------------------------- begin tables -------------------------
\newpage
%-------------------- begin figures ------------------------------------------
\newpage
\clearpage
\end{document}